\documentclass[11pt]{article}
\usepackage{amsfonts,amsmath}
\usepackage{amssymb}
\usepackage{fancyhdr}
\usepackage{slashed}
\usepackage{graphicx}
\usepackage{subfigure}
\usepackage{color}

\usepackage{hyperref}

\usepackage[titletoc]{appendix}

\def\hybrid{\topmargin -20pt    \oddsidemargin 0pt
        \headheight 0pt \headsep 0pt
        \textwidth 6.25in       
        \textheight 9.25in       
        \marginparwidth .875in
        \parskip 5pt plus 1pt   \jot = 1.5ex}

\hybrid

\def\baselinestretch{1.2}

\catcode`\@=11

\def\marginnote#1{}
\newcount\hour
\newcount\minute
\newtoks\amorpm
\hour=\time\divide\hour by60
\minute=\time{\multiply\hour by60 \global\advance\minute by-\hour}
\edef\standardtime{{\ifnum\hour<12 \global\amorpm={am}%
        \else\global\amorpm={pm}\advance\hour by-12 \fi
        \ifnum\hour=0 \hour=12 \fi
        \number\hour:\ifnum\minute<10 0\fi\number\minute\the\amorpm}}
\edef\militarytime{\number\hour:\ifnum\minute<10 0\fi\number\minute}

\def\draftlabel#1{{\@bsphack\if@filesw {\let\thepage\relax
   \xdef\@gtempa{\write\@auxout{\string
      \newlabel{#1}{{\@currentlabel}{\thepage}}}}}\@gtempa
   \if@nobreak \ifvmode\nobreak\fi\fi\fi\@esphack}
        \gdef\@eqnlabel{#1}}
\def\@eqnlabel{}
\def\@vacuum{}
\def\draftmarginnote#1{\marginpar{\raggedright\scriptsize\tt#1}}

\def\draft{\oddsidemargin -.5truein
        \def\@oddfoot{\sl preliminary draft \hfil
        \rm\thepage\hfil\sl\today\quad\militarytime}
        \let\@evenfoot\@oddfoot \overfullrule 3pt
        \let\label=\draftlabel
        \let\marginnote=\draftmarginnote
   \def\@eqnnum{(\theequation)\rlap{\kern\marginparsep\tt\@eqnlabel}%
\global\let\@eqnlabel\@vacuum}  }

\def\preprint{\twocolumn\sloppy\flushbottom\parindent 2em
        \leftmargini 2em\leftmarginv .5em\leftmarginvi .5em
        \oddsidemargin -.5in    \evensidemargin -.5in
        \columnsep .4in \footheight 0pt
        \textwidth 10.in        \topmargin  -.4in
        \headheight 12pt \topskip .4in
        \textheight 6.9in \footskip 0pt
        \def\@oddhead{\thepage\hfil\addtocounter{page}{1}\thepage}
        \let\@evenhead\@oddhead \def\@oddfoot{} \def\@evenfoot{} }

\def\numberbysection{\@addtoreset{equation}{section}
        \def\theequation{\thesection.\arabic{equation}}}

\def\underline#1{\relax\ifmmode\@@underline#1\else
        $\@@underline{\hbox{#1}}$\relax\fi}

\def\titlepage{\@restonecolfalse\if@twocolumn\@restonecoltrue\onecolumn
     \else \newpage \fi \thispagestyle{empty}\c@page\z@
        \def\thefootnote{\fnsymbol{footnote}} }

\def\endtitlepage{\if@restonecol\twocolumn \else \newpage \fi
        \def\thefootnote{\arabic{footnote}}
        \setcounter{footnote}{0}}  

\catcode`@=12
\relax

\def\figcap{\section*{Figure Captions\markboth
        {FIGURECAPTIONS}{FIGURECAPTIONS}}\list
        {Figure \arabic{enumi}:\hfill}{\settowidth\labelwidth{Figure
999:}
        \leftmargin\labelwidth
        \advance\leftmargin\labelsep\usecounter{enumi}}}
 \relax
\def\tablecap{\section*{Table Captions\markboth
        {TABLECAPTIONS}{TABLECAPTIONS}}\list
        {Table \arabic{enumi}:\hfill}{\settowidth\labelwidth{Table
999:}
        \leftmargin\labelwidth
        \advance\leftmargin\labelsep\usecounter{enumi}}}
 \relax
\def\reflist{\section*{References\markboth
        {REFLIST}{REFLIST}}\list
        {[\arabic{enumi}]\hfill}{\settowidth\labelwidth{[999]}
        \leftmargin\labelwidth
        \advance\leftmargin\labelsep\usecounter{enumi}}}
 \relax
\makeatletter
\newcounter{pubctr}
\def\publist{\@ifnextchar[{\@publist}{\@@publist}}
\def\@publist[#1]{\list
        {[\arabic{pubctr}]\hfill}{\settowidth\labelwidth{[999]}
        \leftmargin\labelwidth
        \advance\leftmargin\labelsep
        \@nmbrlisttrue\def\@listctr{pubctr}
        \setcounter{pubctr}{#1}\addtocounter{pubctr}{-1}}}
\def\@@publist{\list
        {[\arabic{pubctr}]\hfill}{\settowidth\labelwidth{[999]}
        \leftmargin\labelwidth
        \advance\leftmargin\labelsep
        \@nmbrlisttrue\def\@listctr{pubctr}}}
 \relax
\makeatother
\newskip\humongous \humongous=0pt plus 1000pt minus 1000pt

\newif\ifdtup

\relax

\def\be{\begin{equation}}
\def\ee{\end{equation}}
\def\ba{\begin{eqnarray}}
\def\ea{\end{eqnarray}}

\def\a{\alpha}

\def\g{\gamma}
\def\G{\Gamma}

\def\e{\epsilon}

\def\m{\mu}

\def\l{\lambda}

\def\s{\sigma}

\def\cR{{\cal R}}

  \def\cC{{\cal C}}

  \def\cO{{\cal O}}
\def\cP{{\cal P}}  \def\cR{{\cal R}}
 \def\cT{{\cal T}}

\newcommand{\vev}[1]{{\left< {#1} \right>}}

\newcommand{\prt}[1]{{\left( {#1} \right)}}
\newcommand{\prtt}[1]{{\left[ {#1} \right]}}

\def\no{\noindent}

\def\IR{\relax{\rm I\kern-.18em R}}

\def\pp{\partial}

\def\IR{\relax{\rm I\kern-.18em R}}
\def\IL{\relax{\rm I\kern-.18em L}}

\def\inv{^{\raise.15ex\hbox{${\scriptscriptstyle -}$}\kern-.05em 1}}

\def\cR{{\cal R}}

\def\bea{\begin{eqnarray}}
\def\eea{\end{eqnarray}}
\newcommand{\eq}[1]{(\ref{#1})}

\newcommand{\la}[1]{\label{#1}}

\def\a{\alpha}      
       
\def\g{\gamma}  \def\G{\Gamma}  
    
\def\e{\epsilon}

\def\l{\lambda} 
\def\m{\mu}

\def\s{\sigma}

\definecolor{markcolor2}{rgb}{1,0,0}

\definecolor{markcolor3}{rgb}{0,1,0}

\newcommand{\ff}{\frac}

\begin{document}

\renewcommand{\theequation}{\thesection.\arabic{equation}}
\csname @addtoreset\endcsname{equation}{section}

\newcommand{\beq}{\begin{equation}}
\newcommand{\eeq}[1]{\label{#1}\end{equation}}
\newcommand{\ber}{\begin{eqnarray}}
\newcommand{\eer}[1]{\label{#1}\end{eqnarray}}
\newcommand{\eqn}[1]{(\ref{#1})}
\begin{titlepage}

\begin{center}

~
\vskip 1 cm

{\Large
\bf Holographic Observables at Large $d$}

\vskip 0.8in

{\bf Dimitrios Giataganas$^{1,2}$ ,\phantom{x} Nikolaos Pappas$^{3}$, \phantom{x} Nicolaos Toumbas$^{4}$}
\vskip 0.2in
{\em
  ${}^1$  Department of Physics, National Sun Yat-sen University,  \\
Kaohsiung 80424, Taiwan
\vskip .1in
 ${}^2$  Center for Theoretical and Computational Physics,   \\
Kaohsiung 80424, Taiwan
\vskip .1in
${}^3$ Department of Physics, University of Athens, \\
University Campus, Zographou 157 84, Greece
\vskip .1in
${}^4$
Department of Physics, University of Cyprus,\\
Nicosia 1678, Cyprus
\\\vskip .15in
{\tt dimitrios.giataganas@mail.nsysu.edu.tw}, {\tt pappasnikolaos.uni@gmail.com},  {\tt nick@ucy.ac.cy}
}

\vskip .4in
\end{center}

\vskip .4in

\centerline{\bf Abstract}
We study holographically non-local observables in field theories at finite temperature and in the large $d$ limit. These include the Wilson loop, the entanglement entropy, as well as an extension to various dual extremal surfaces of arbitrary codimension. The large $d$ limit creates a localized potential in the near horizon regime resulting in a simplification of the analysis for the non-local observables, while at the same time retaining their qualitative physical properties. Moreover, we study the monotonicity of the coefficient $\a$ of the entanglement's area term, the so called area theorem. We find that the difference between the $UV$ and $IR$ of the $\a$-values, normalized with the thermal entropy, converges at large $d$ to a constant value which is obtained analytically. Therefore, the large $d$ limit may be used as a tool for the study and (in)validation of the  renormalization group monotonicity theorems.  All the expectation values of the observables under study show rapid convergence to certain values as $d$ increases. The extrapolation of the large $d$ limit to low and intermediate dimensions shows  good quantitative agreement with the numerical analysis of the observables. \no
\end{titlepage}
\vfill
\eject


\noindent


\def\baselinestretch{1.2}
\baselineskip 19 pt
\noindent


\setcounter{equation}{0}

\section{Introduction}

The non-local observables have played an important role in the study of theoretical physics over the years. The gauge/gravity duality \cite{Aharony:1999ti,CasalderreySolana:2011us} has provided a solid tool for their study at strong coupling, with several interesting applications. Non-locality in the context of gauge/gravity duality is in principle associated to minimization problems in curved spacetimes. These are mainly extremal surfaces, which give rise to different non-local observables according to their dimensionality. Such minimization problems are very challenging even on flat spaces, and one would expect that they would be unsolvable in curved space-times. Many of them are; however in several cases the symmetry of the spacetime and the simplicity of the boundary conditions make the problem tractable.

Wilson loop and entanglement entropy are two of these prominent non-local observables. The Wilson
loop exponent is proportional to the leading contribution to the effective potential between two static quarks and serves as an order parameter for confinement. The entanglement entropy is the von Neumann entropy of the reduced density matrix of a partition of the Hilbert space.  The Wilson loop corresponds to a two-dimensional minimal surface hanged from the boundary of the holographic space, while the entanglement entropy to a spatially co-dimension one surface. These type of surfaces can have strip boundaries that exhibit translational invariance on the directions transverse to the strip separation. When the theory is conformal their regularized areas can be found in closed form in terms of elementary functions. The regularization of the ultraviolet contributions is necessary for any such observable. In entanglement these arise from local UV physics near the entangling surface, while for the Wilson loop they amount to the contribution to the interquark potential of the infinite mass of the fundamental quarks. From the minimal surface perspective, the UV contributions are associated with the infinite distance required to reach the boundary from the bulk.

In the presence of scales the non-local observables reveal the richer properties of the theory but do not acquire a closed form in terms of the physical scales of the theory.  At finite temperature, as the separation of the quarks increases, the Wilson loop undergoes a phase transition from a connected surface to a pair of disconnected surfaces, which become energetically favorable. The phase transition depends on the possible scales of the theory and occurs at an intermediate temperature, where its analytical details are not fully tractable. The holographic entanglement entropy at zero temperature and even dimensions is related to the central charge coefficient of the Euler density contribution to the conformal anomaly, while in odd dimensions a connection to the F-theorems has been shown
\cite{cardyc,Giombi:2014xxa}. This has motivated the proposal of several c-function candidates whose monotonicity is tighten up to the satisfaction of the null energy conditions \cite{Ryu:2006ef,Myers:2010tj,Chu:2019uoh,Giataganas:2017koz} with interesting applications in holography \cite{Baggioli:2020cld,Cremonini:2020rdx,Hoyos:2021vhl,Cartwright:2021hpv,Arefeva:2020uec,Cremonini:2013ipa}. Another monotonic function along the Renormalization Group (RG) flow is defined by the area theorem.  For entangling regions that cover almost the entire spacetime, the entanglement entropy approaches the thermal entropy exhibiting area subleading contributions. The coefficient of the area $\a$ has been found to be decreasing along Lorentz invariant RG flows \cite{Casini:2012ei,Casini:2016udt}. The entanglement entropy as well as the holographic functions $\a$ and $c$ have not been expressed in an exact closed form in the presence of scales. Moreover, their monotonicity is guaranteed only for theories with Lorentz symmetry and therefore it is an interesting task to analytically obtain them and study their properties along the different RG flows.

In our approach we study these phenomena in the limit of large number of dimensions $d$, both analytically and numerically. We rely on the fact that the bulk geometry is well defined at any dimension and we will assume that the dictionary between extremal surfaces and non-local observables is valid at any $d$. Moreover, the large $d$ expansion in our holographic study can be seen as a conceptual and computational tool, since all the computational limits are smooth. Although we note that the gauge/gravity  correspondence is microscopically well-defined only for low dimensions and interacting conformal field theories have been known only  for $d\le 6$, there is no obstruction in defining the geometry at large $d$ and assuming that holography holds there.  The computation of holographic observables at large $d$ can be always done given the aforementioned assumptions and the extrapolation to lower dimensions is always conceptually and computationally possible, since smooth limits are involved. The story carries some distant similarities to the large-$N$ expansion in gauge theories. A motivation for our study is to examine whether at large $d$ the geometrical computations are simplified. At the same time one may study whether certain qualitative properties of theories at finite $d$ are captured by the large $d$ expansion, and how well the quantitative ones can be approximated by the extrapolation of the large $d$ results. Notice that there are several interesting developments in general relativity at large $d$, which also serve as a motivation, see for example \cite{Emparan:2013moa,Emparan:2013xia,Emparan:2014cia,Dandekar:2016fvw}. Surprisingly, the large $d$ expansions have been proven to be both qualitative and in many cases even quantitatively accurate for the gravitational studies, resembling in some sense the behavior of the large-$N$ extrapolation to lower values in gauge theories. The explanation relies partly on the fact that for large $d$, the gravitational potential becomes extremely steep, which tends to separate the dynamics of the near horizon region and the rest of the space, while preserving most of the characteristics of the gravitational potential. We will see that this is the case for the holographic studies in this paper.

 The large $d$ expansion has been already used successfully in the context of holography in computing the zero temperature phase transitions of holographic mutual information \cite{Colin-Ellerin:2019vst}, in the study of the holographic momentum relaxation and of the superconductors \cite{Andrade:2015hpa,Garcia-Garcia:2015emb}, and in hydrodynamic related studies \cite{Rozali:2017bll,Casalderrey-Solana:2018uag,Andrade:2018zeb}. Here we apply the large $d$ expansion in holography for the computation of non-local observables. The expectation value of the observable is determined  by the boundary data of the problem, in this case the length $L$ of the surface, and the rest of the scales of the theory. The main difficulty lies in expressing the area of the surface in the bulk in terms of $L$ and the rest of the scales in the theory. There have been numerous interesting approaches to tackle this problem, including mainly those of applying different type of series on the surfaces  \cite{Brandhuber:1998bs,Fischler:2012ca,Erdmenger:2017pfh}.

In our work we study how Wilson loops and Wilson surfaces of arbitrary codimension behave in the large $d$ limit in thermal theories and in theories that exhibit confinement. We find that the analysis becomes simpler in this limit, while all the qualitative features of the theory concerning the observables are still captured. Moreover, quantitative extrapolation at low $d$ shows a good agreement in many cases. Then we move on to study entangling surfaces at large $d$. We find again that the limit reduces the complexity of the analytical computations. In addition, we compute the coefficient $\a$ of the so-called area theorem that has been found to exhibit monotonicity properties along the RG flow. We find that the difference of the UV and the IR contributions to the function $\a$ converges in the large $d$ limit to a certain number, simplifying tremendously the computation. This number reveals the violation of the area theorem in the theories under discussion, as expected, due to the breaking of Lorentz invariance. This suggests that the large $d$-limit is an invaluable tool for the (in)validation of the  RG monotonicity theorems. We also study spatial surfaces of arbitrary codimension  with similar profile as the entangling surfaces and derive their properties at large $d$. We note that all the expectation values of the observables under study show rapid convergence to certain values as $d$ increases. We observe that the extrapolation of the large $d$ limit  to low dimensions shows a good quantitative agreement with the numerical analysis on the observables.

\section{Gravity Backgrounds}

We consider spherically symmetric black holes in $AdS_{d+1}$ spacetimes.  Small $AdS$ black holes exhibit similar behavior with asymptotically flat black holes, which are unstable. Large $AdS$ black holes have positive specific heat and are thermodynamically stable. They describe the high temperature phase of the dual field theory. For large mass they become planar with a translationally invariant horizon
\be\la{met11}
ds^2_{d+1}=-\ff{r^2}{R^2} f(r)  dt^2+\ff{R^2 dr^2}{r^2 f(r)}+r^2 dx_{d-1}^2~,\qquad f(r):=\prt{1-\ff{r_{h}^{d}}{r^{d}}}~.
 \ee
In order to avoid a conical singularity at the horizon, the temperature must be
\be\la{ttt}
T=\ff{d r_h}{4 \pi R^2}~.
\ee
The dual boundary theory lives on $R^d$. It is  strongly coupled and has temperature $T$.

The following complex coordinate transformation \cite{Horowitz:1998ha,Hubeny:2009rc}
\be\la{ccord}
t= 2 \pi i T_{pl} R^2 \phi_{pl}~,\qquad \phi=  -2 \pi i T_{pl} t_{pl}~,\qquad r=\ff{r_{pl}}{2 \pi T_{pl} R} ~,\qquad x=2 \pi T_{pl} R x_{pl}~,\qquad r_k=r_h~,
\ee
relates the planar black hole to the soliton solution. The subscript $pl$ refers to quantities associated with metric \eq{met11}.  The soliton metric reads
\be\la{soliton}
ds_{soliton}^2{}_{d+1}=\ff{r^2}{R^2}\prt{-dt^2+\prt{1-\ff{r_k^{d}}{r^{d}}} R^2 d\phi^2}+r^2 dx_{d-2}^2+\ff{R^2 }{r^2\prt{1-\ff{r_k^{d}}{r^{d}}}} dr^2~.
\ee
Notice that the IR tip of the geometry is at $r_k$ and this is the reason that the boundary theory confines.

There are several ways to define the large $d$ limit. A convenient way is to keep the horizon radius $r_h$ fixed.  As we increase the number of dimensions, the potential becomes more and more localized around the horizon of the black hole. In the near horizon regime, the potential develops a large gradient. For $d\rightarrow \infty$ the observables in the geometry are approaching closer to the zero temperature ones. The large $d$ expansion however has crucial differences compared to the zero temperature results and preserves the finite temperature  properties of the theory. The low temperature expansion corresponds to having the black hole horizon far away from the boundary. On the other hand, in the large $d$ expansion the black hole horizon can be close to the boundary and the expansion relies on the localization of the potential around the horizon.

Moreover, notice that the limit of large $d$ with $r_h$ fixed leads to an effectively large temperature $T$. Qualitatively, the observables in this limit will reflect upon a hotter environment and therefore they will signal phenomena associated with high temperature. For example, as $d$ increases, we expect the critical length $L_c$ associated with the Wilson loop phase transition to decrease. The exact analysis of how this happens and the effect of the large $d$ expansion is one of the subjects of this paper.

In what follows we will study the holographic non-local observables in the large $d$ limit for the  backgrounds \eq{met11} and/or \eq{soliton}.

\section{Basics of Wilson Loops and large $d$}

One of the most interesting observables in strongly coupled gauge theories is the Wilson loop, which acts as an order parameter of confinement. It is given by the path-ordered exponential of the gauge field traced in the fundamental representation
\be\la{wl1}
W_\cC=\ff{1}{N}Tr \cP exp\prt{i\oint A_\m dx^\m}~.
\ee
It is a non-local observable where the expression above can be integrated along any path in space. Its physical interpretation is that taking an infinitely massive quark in the fundamental representation along a loop, it will be transformed by the factor \eq{wl1}. In this sense the Wilson loop exponent is proportional to the leading contribution to the potential between the quarks in a heavy meson state. Its expectation value can be expressed in terms of the energy eigenstates of the corresponding Hamiltonian, where in the limit of large time $\cT$, the dominant contribution comes as follows
\be
\vev{W_\cC}\simeq e^{-V(L) \cT},\qquad \cT\rightarrow \infty ~.
\ee
In gauge/gravity duality the expectation value of the Wilson loop is given by the action of a minimal surface bounded by the curve on the boundary of space.  Its expectation value is given by
\be
\vev{W_\cC}\simeq e^{\sqrt{\l }A}~,
\ee
where $A$ is the regularized area of the minimal surface in the curved space bounded by the boundary loop $\cC$.

The question here is how the number of dimensions affects the Wilson loop expectation value. In holography the action of the minimal surface with a rectangular boundary consisting of a spatial edge of length $L$, which represents the  distance of the heavy static quarks along the $x$ direction, and the time direction of length $\cT\rightarrow \infty$ is given by
\be
S=\ff{\cT}{2\pi \a'} \int d\s \sqrt{-g_{tt} \prt{g_{rr}r'{}^2+g_{xx}}}~.
\ee
We have considered a holographic element of the form $ds^2=g_{tt}dt^2+g_{xx}d\vec{x}^2+g_{rr} dr^2$, where the functions depend on the holographic coordinate $r$. The boundary of space is at $r=\infty$ where the spacetime metric elements $g_{tt}$ and $g_{xx}$ diverge. The equation of the surface in the bulk is given by integrating with respect to the loop parameter the following expression
\be\la{equ}
 r'(\s)^2=-\ff{g_{xx} \prt{g_{tt} g_{xx}+c^2}}{c^2 g_{rr}}~,\qquad c^2:=-g_{tt} g_{xx}\big|_{r_0}~,
\ee
where $r_0$ is the turning point of the surface in the bulk. The latter is related to the length of the surface on the boundary. There are several ways the large $d$ limit may be taken. Here, we increase the number of dimensions of the theory while we keep $r_h$ fixed. As $d$ increases, an increasingly larger, more localized potential appears around the horizon of the black hole in the bulk.  For $d\rightarrow \infty$ the solution of the equations is expected to approach the zero temperature minimal surface. The large $d$ expansion however has crucial differences compared to the zero temperature result and preserves the properties of the finite temperature solutions. For each value of the turning point $r_0$, equation \eq{eqld} gives two solutions with different boundary values, a characteristic of finite temperature 2-dimensional minimal surfaces.

The boundary length $L$ of a Wilson loop corresponds to the size of a heavy meson extended along the direction $x$ and is given by the integration of \eq{equ} as
\be\la{lwl}
L= 2 \int_{r_0}^{r_b} dr \sqrt{\frac{-g_{rr} c^2}{g_{xx} \prt{g_{tt} g_{xx}+c^2}}}~,
\ee
where $r_b$ is the $r$-value at the boundary of the theory. The dimension of the space $d$ enters in the expressions above only through the metric elements since the surface is of fixed dimensionality. The area of the minimal surface can be also expressed in terms of the turning point $r_0$. It is always divergent due to the infinite distance of the bulk from the boundary in holographic space-times. To regularize the infinities, we introduce a radial cut-off $r_b$ which is large but finite and add a counterterm $S_c$ that renormalizes $S$:
\be\la{stot}
S=S_A-S_{c}~,
\ee
where
\be\la{sat1}
S_A=\ff{\cT}{\pi \a'}\int_{r_0}^{r_b} dr \sqrt{-\ff{g_{xx} g_{rr} g_{tt}^2}{g_{tt} g_{xx}+c^2}}~.
\ee
At the end we take the limit $r_d\rightarrow \infty$. 
Notice that the integral above has been doubled to take into account the whole string by incorporating the symmetry of the string. The integral along the radial dimension depends on the number of the dimensions through the metric functions.

\subsection{Renormalization of Infinities and their Dependence on $d$}

The renormalization of  infinities can be done in several ways. The infinite distance from the boundary corresponds to the infinite mass of the meson quark and antiquark. So subtracting their masses, it is their interaction energy that remains. In fact the counterterm $S_c$ in \eq{stot} is given by the dominant $r_b$ dependent term in the expression of the infinite mass of the quarks given by
\be\la{sren}
S_{m}=\ff{\cT}{\pi \a'}\int_{r_k}^{r_b} dr \sqrt{-g_{tt} g_{rr} }~,
\ee
where $r_k$ is the deepest point in the bulk that a static straight string originating from a point at the boundary can reach. The counterterm depends
only on intrinsic variables of the cutoff surface at $r_b$, while $S_m$ depends also on the sate of the theory.
In the presence of horizons or cigar-type geometries, $r_k$ denotes the horizon or the position of the tip of space respectively. Therefore, the dependence on the dimensions $d$ potentially enters through the metric elements in this expression, although this is not necessary, since the product in the integrand can be $d$-independent.

The independence of the UV divergence on $d$ in certain theories is not surprising. The divergence is related to the way that the string approaches the boundary. It is perpendicular to the spatial dimensions running along the $r$-dimension, and therefore, for symmetric spaces like the planar $AdS$ black holes, there should be no $d$-dependence, as indeed happens.

An alternative renormalization scheme is motivated by the different type of boundary conditions that the string has in the $d$-dimensional spacetime. Let us consider a theory in $d_1>d$ dimensions in the presence of a number of space filling $(d_1-1)$-dimensional branes. The Wilson loop in this case can be thought of as corresponding to an open string bounded by the loop $\cC$ with Dirichlet boundary conditions, since the string endpoint has complementary Neumann boundary conditions along the space filling $d_1-1$ branes. To reduce the theory in $d$-dimensions we perform T-dualities along the $d_1-d$ directions, which transform the Dirichlet boundary conditions to Neumann ones, to end up with a Wilson loop obeying $d$ Dirichlet boundary conditions and $d_1-d$ Neumann boundary conditions.

The above discussion motivates the Legendre transform \cite{Drukker:1999zq,Chu:2008xg}, since $\vev{W}$ should be thought of as a functional of the coordinates in $d$ dimensions and the momenta in $d_1-d$ dimensions, which in our discussion here only the radial one plays the major role. The Legendre transform is not diffeomorphism invariant, so we work in the Poincare-like set of coordinates where it reads
\be
S_{Leg}=\ff{\cT}{\pi \a'} \int_{0}^{L/2} dx~ p_r r'=\ff{\cT}{\pi \a'} \prt{p_r ~r}\bigg|_{r_0}^{r_b}~,
\ee
where $p_r$ is the conjugate momentum to the $r$ coordinate and does not depend explicitly on $x$.
From this one may read the counterterm
\be\la{sa1}
S_{~ c}\simeq   \ff{\cT}{\pi \a'}   \prt{r \sqrt{-g_{tt} g_{rr}}}\bigg|_{r=r_b}~,
\ee
where at the boundary the term $g_{tt} g_{rr}$ diverges and the expression depends
only on intrinsic variables of the cutoff surface at $r_b$.
Note that the Wilson loops in the adjoint representations are dual to higher dimensional branes that have similar profiles to the strings discussed above. The renormalization of the infinities applies in a similar manner. Notice that from now on we will drop the overall $\cT$ factor to simplify the presentation.

\section{Holographic Wilson Loop at Large $d$}

The Wilson loop surface at zero temperature is unique for fixed boundary conditions. At finite temperature there are two extremal surfaces with the same boundary. One extremal surface, $S_{unstable}$, is the unstable one that goes deep into the holographic direction and has area $S_{unstable}\ge S_{stable}$. $S_{stable}$ is the stable surface that remains closer to the boundary. It is the one that is physically relevant for the meson potential. In the stable surface brunch there is one more competition of dominance among the different solutions. $S_{stable}$ is the connected surface between the two end-points, which competes in terms of area or energy with the two disconnected straight-line surfaces $S_{m}$. The Wilson loop phase transition takes place when $S_{stable}=S_{m}$, which, for a fixed length, occurs with the  increase of the temperature of the theory. This could be interpreted as the way the meson melts down to its constituents, since the bound state is not anymore dominant compared to having the quarks separated.

To extract the minimal surface dependence on $d$, we substitute the metric elements \eq{met11} in the Euler-Lagrange equation \eq{equ}. Writing explicitly this equation for the holographic background under study, we get
\be\la{eqld}
r'(\s)^2=-f(r) r^4 \prt{1-\ff{f(r) r^4}{f(r_0)r_0^4}}~.
\ee
In this section we set $R=1$. For $d\rightarrow \infty$ the solution to the equations approaches the zero temperature minimal surface. The large $d$ expansion of the finite temperature theory however has crucial differences compared to the zero temperature theory. For example, for each value of the turning point $r_0$,  equation \eq{eqld} gives two solutions with different boundary values. 
Equation \eq{lwl} gives for the boundary length
\be\la{llll}
L=2\int_{r_0}^{\infty} dr \ff{1}{r^2}\ff{1}{\sqrt{f(r)\prt{\ff{r^4f(r)}{r_0^4 {f(r_0)}}-1}}}:=\int_{r_0}^{\infty} dr L_{int}~.
\ee
The surfaces that are close to the boundary of the space and get deeper in the bulk as $r_0$ decreases, will have increasing boundary distance $L$. On the other hand, for surfaces that are very deep into the bulk in the near horizon regime, as $r_0$ decreases, their boundary distance $L$ decreases. Therefore,   $L(r_0)$ is a function with a maximum point for some $r_0$, irrespective of the number of dimensions. The dependence of the saddle point on the number of dimensions can be estimated by an expansion of the derivative of the integrand around $r\sim r_0$.
For $d>4$, the value of $r_0$ at the maximum as a function of $d$, can be approximated to be
\be
r_0{}_{max}\simeq16^{-1/d} (d-4)^\ff{2}{d}{\,r_h}~,
\ee
by setting $\ff{d L_{int}}{ d r_0}$ equal to zero and obtaining the leading contributions near the boundary. Especially, for large $d$, the approximation improves rapidly and reads
\be
r_0{}_{max}\simeq r_h\prt{1+\ff{2}{d}\log{\ff{d}{4}}+\ff{2}{d^2}\prt{\prt{\log\ff{4}{d}}^2-4}}~.
\ee
This approximated expression captures well the qualitative behavior of the dependence of minimal surfaces on the number of dimensions. For low dimensions, as $d$ increases, the maximum of $L(r_0)$ moves towards the bulk, until a critical dimension is reached, still in the regime of low $d$, where the monotonicity changes. Further increase of $d$ moves the maximum closer to the boundary. 

This summarizes the qualitative behavior of the minimal surfaces and how, in the large $d$ limit, they approach the zero temperature surfaces. This analytical treatment is justified by the numerical analysis of the latter sections and is summarized in Figure \ref{figure:Lr0t}. In the numerical sections we elaborate more on the features of the minimal surfaces and their dependence on the number of dimensions $d$ presenting the accurate analysis.

A further analytical treatment involves an expansion of the quantities in terms of $\e=1-r_h^d/r_0^d$. Expansion of \eq{llll} with respect to $\e$ and integration gives
\be\la{expe1}
r_0\simeq \ff{\sqrt{\pi}}{L}\prt{\ff{2\Gamma\prt{\ff{7}{4}}}{3\Gamma\prt{\ff{5}{4}}}+\ff{\e-1}{24 }\prt{\ff{3 (d-1)\Gamma\prt{\ff{3+d}{4}}}{\Gamma\prt{\ff{5+d}{4}}}-\ff{16~ \Gamma\prt{\ff{7}{4}}}{\Gamma\prt{\ff{1}{4}}} }}~.
\ee
On the other hand, to probe the phase transition we compute the quantity $S_A-S_{c}-\prt{S_m-S_c}$ which eventually leads to
\be \la{sa12}
S_A-S_{m}= -\ff{r_0}{\sqrt{\pi}\a'}\prt{\ff{\Gamma\prt{\ff{7}{4}}}{3\Gamma\prt{\ff{5}{4}}}+ \ff{\e-1}{12}\prt{\ff{4 \Gamma\prt{\ff{7}{4}}}{\Gamma\prt{\ff{5}{4}}}-\ff{3 \Gamma\prt{\ff{3+d}{4}}}{\Gamma\prt{\ff{1+d}{4}}} }}+\ff{r_h}{\pi\a'}~,
\ee
where from \eq{sren}
\be
S_{m}=\ff{1}{\pi\a'}\int_{r_h}^{r_b} d r~.
\ee
This quantity is equal to the interaction potential between the quark-antiquark pair, which is obtained by subtracting from the energy the infinite quark mass  (including the thermal contribution).
The equation written in this form reveals the special properties of the $d=4$ case, where the $\Gamma$ functions combine and simplify considerably.  Equation \eq{sa12} has an immediate consequence on the melting of the meson at large $d$.  The mass term depends inversely proportional on $d$ only through the horizon of the black hole \eq{ttt}. By keeping fixed the horizon position as we increase the dimensionality of the theory, the mass term remains the same. The melting of the meson occurs for the $L$ value for which the potential becomes equal to the $S_m$. Assuming that the form of the potential almost saturates for some value of $d$, we expect for large $d$, $L_c \simeq c_1 $, which implies that
\be\la{erl}
L_c T\sim c_1 d + \cO(d)~.
\ee
Indeed this is what we confirm with our numerical analysis in the next section, where we also determine the subleading terms. The assumption made on a converging form of the potential at large enough values of $d$ is natural, since the gravitational potential localizes in the near horizon regime.

We can also try to determine the constant $c_1$. By combining equations \eq{sa12} and \eq{expe1} we obtain
\be\la{stota1}
S_{tot}=S_A-S_m=-\ff{1}{2\a'}\ff{\Gamma\prt{\ff{7}{4}}}{ \Gamma\prt{\ff{5}{4}} L} \prt{\ff{4 \Gamma\prt{\ff{7}{4}}}{9\Gamma\prt{\ff{5}{4}}}+\ff{r_h^d}{r_0^d}\ff{\Gamma\prt{\ff{3+d}{4}}}{6 \Gamma\prt{\ff{5+d}{4}}}}+\ff{r_h}{\pi\a'}~,
\ee
where the conformal contribution $L^{-1}$ separates as we take the large $d$ limit.  The expression that determines the critical length is given implicitly at large $d$ as
\be\la{lcbc}
L_c\simeq \ff{ \pi \Gamma\prt{\ff{7}{4}}}{3\Gamma\prt{\ff{5}{4}} r_h }\prt{\ff{ 2\Gamma\prt{\ff{7}{4}}}{3\Gamma\prt{\ff{5}{4}}}+
\ff{1}{2 \sqrt{d}}\ff{r_h^d}{r_0^d}}~.
\ee
Although it cannot be solved analytically to give $L_c(T)$, we may confirm that as $d$ increases, the critical length decreases, as long as we keep $r_h$ fixed, and
the large $d$ result \eq{erl} holds. In particular \footnote{One may use the properties of $\Gamma$ functions, $\Gamma(1+z)= z \Gamma(z)$ and  $\Gamma(1-z)=\pi/\prt{\sin (\pi z) \Gamma(z)}$, to rewrite the expressions in equivalent forms. For example, equation \eq{lcbc} is written in terms of $\Gamma(1/4)$.}
\be\la{erl2}
L_c T\simeq  \ff{  \pi^2}{\Gamma\prt{\ff{1}{4}}^4   } d+\ldots~.
\ee
Notice that the expansion resembles the low $T$ expansion, although here all the expressions are valid for high temperatures (in contrast to the low $T$ expansion). This is why we are allowed to determine approximately the critical length.  We will confirm, in the numerical section, that the extrapolation to low and intermediate values of $d$ is well justified.

\subsection{Large $d$ Numerics}

Having presented the analytical treatment of the Wilson Loops at large $d$, we now turn to compute the parameters numerically for intermediate and large number of dimensions, where the analytical calculations and particularly the evaluation of the integrals are intractable.

We begin the numerical integration by observing the behavior of a minimal surface of a fixed turning point in the bulk as we change the dimensionality of the theory. The question we ask is how the extremal surface on the boundary will adapt as we increase the number of dimensions in order to keep fixed the turning point $r_h/r_0$. Ideally, we would like to find a quick converging behavior at large $d$, as we have argued in the analytical study.

\begin{figure}
\begin{minipage}[ht]{0.5\textwidth}
\begin{flushleft}
\centerline{\includegraphics[width=75mm]{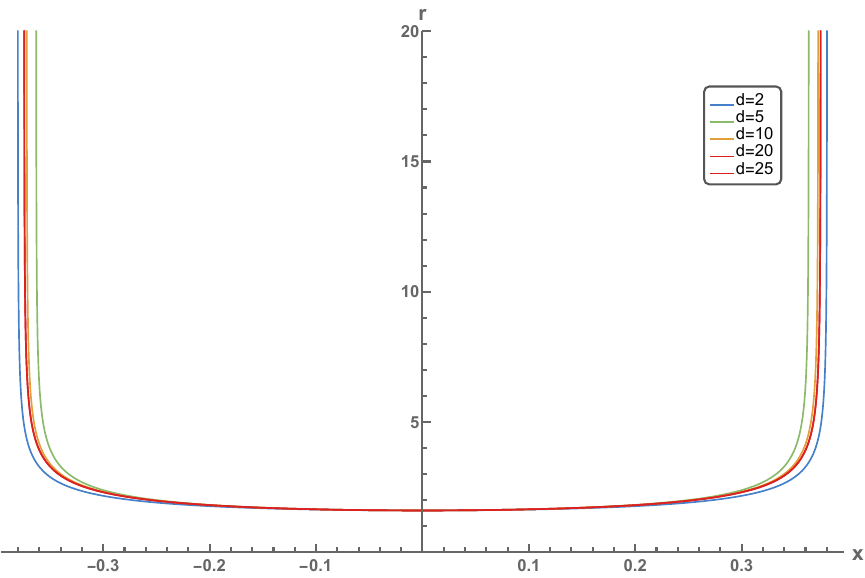}}
\caption{\small{The shape of the stable branch surfaces with fixed turning point $r_0=1.6 r_h$. For the energetically favorable surfaces there is a quick convergence to a certain surface, as we increase the number of dimensions.}}
\label{figure:rd1}
\end{flushleft}
\end{minipage}
\hspace{0.3cm}
\begin{minipage}[ht]{0.5\textwidth}
\begin{flushleft}
\centerline{\includegraphics[width=75mm ]{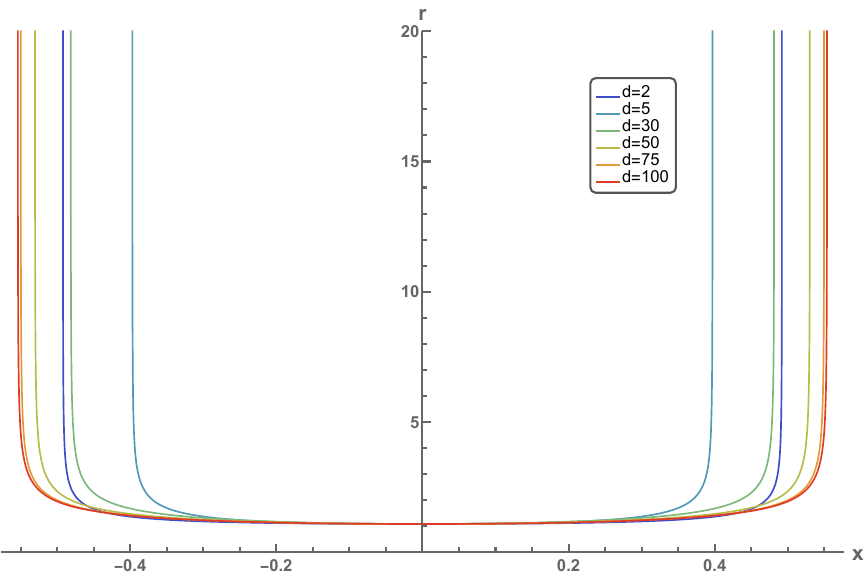}}
\caption{\small{The shape of the energetically non-favorable surfaces with $r_0=1.08 r_h$, probing regimes close to the black hole horizon. There is still a convergence to a certain surface but it is slower compared to the stable surfaces.}}
\label{figure:rd2} \vspace{.0cm}
\end{flushleft}
\end{minipage}
\end{figure}

\begin{figure}
\begin{minipage}[ht]{0.5\textwidth}
\begin{flushleft}
\centerline{\includegraphics[width=75mm]{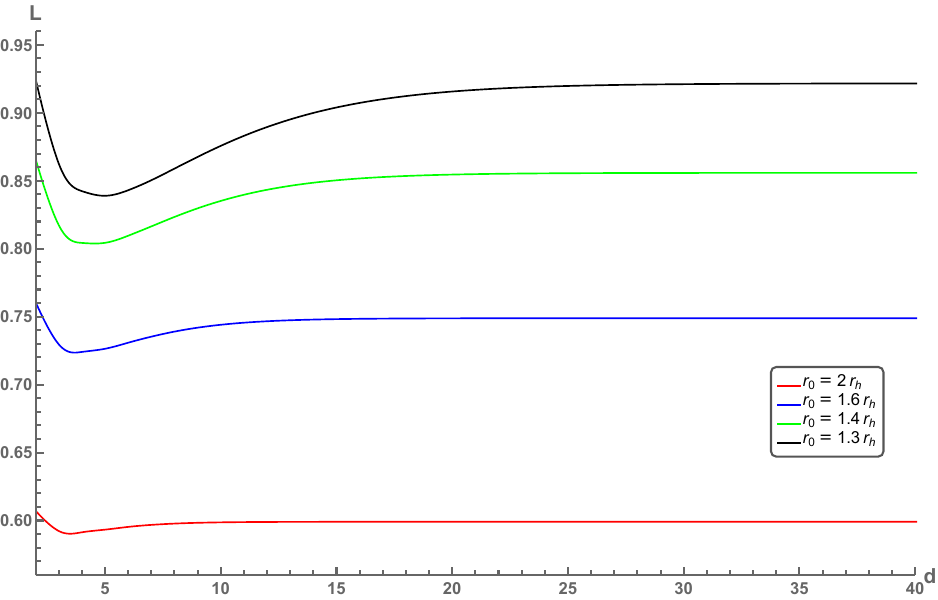}}
\caption{\small{The boundary length for surfaces with fixed turning point with respect to the dimension of space. The turning point is on the stable branch of the surface and it is the one that is physically relevant. We observe a mild non-monotonic behavior for low-intermediate $d$ and a quick convergence on a certain length $L$ with increasing number of dimensions.}}
\label{figure:Lr1}
\end{flushleft}
\end{minipage}
\hspace{0.3cm}
\begin{minipage}[ht]{0.5\textwidth}
\begin{flushleft}
\centerline{\includegraphics[width=75mm ]{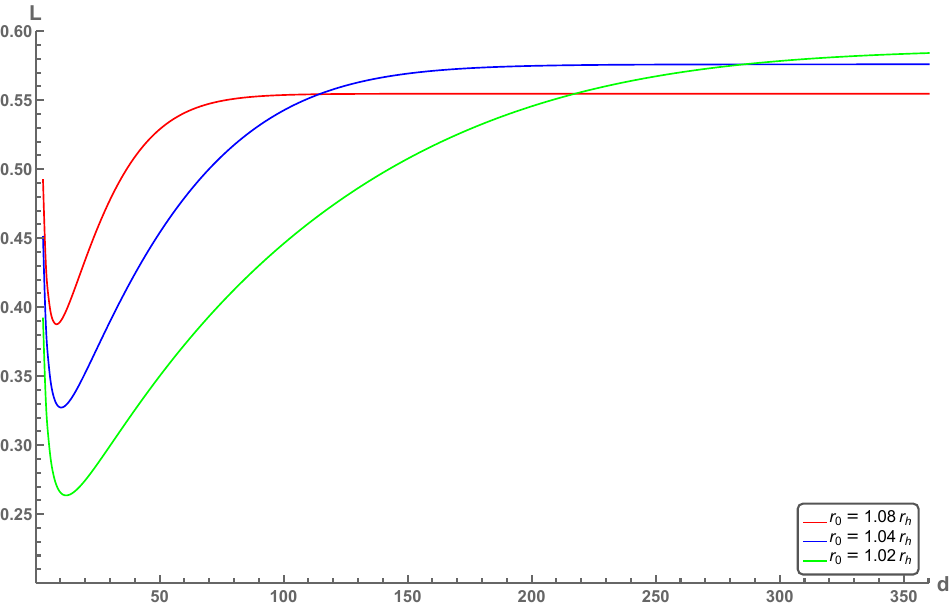}}
\caption{\small{The boundary length for surfaces with fixed turning point on the unstable branch. These surfaces probe the regime close to the black hole horizon. We observe a clear non-monotonic behavior for low-intermediate values of $d$ and a slower convergence on a certain length $L$ as the number of dimensions increases. This is expected since these surfaces are closer to the localized near horizon potential.}}
\label{figure:Lr2} \vspace{.0cm}
\end{flushleft}
\end{minipage}
\end{figure}

For surfaces  with turning points close to the horizon of the planar black hole, we observe that the boundary length develops a slight initial decrease with increasing $d$, while still for low $d$, the monotonicity changes and $L$ increases until a saturation point. These surfaces belong to the unstable branch of the extremization problem.  For the physically interesting surfaces in the stable branch that do not probe the near-horizon regime, there is again a slight decrease of $L$, while for a certain value of $d$, $L$ starts increasing with $d$ until a saturation is observed. We find that the saturation of the boundary length $L$ happens at values that can be considered to be of low dimensionality, especially for the surfaces in the stable branch. This can be understood from the fact that the geometry closer to the boundary becomes increasingly insensitive to the number of dimensions as $d$ increases, and therefore, the minimal surfaces that probe these regimes will reach faster  the convergent length $L$. The overall behavior of the physical branch of the surfaces is  plotted in  figure \ref{figure:rd1}, and for the unstable surfaces in  figure \ref{figure:rd2}, where we plot the minimal surfaces. Moreover, in figures \ref{figure:Lr1} and \ref{figure:Lr2} we plot the boundary length of the surfaces, corresponding to the size of meson, for various fixed turning points in the stable and unstable branch respectively.

The purpose of the study so far is to examine the geometric properties of the surfaces which generate the strongly coupled physics. For this reason and for presentation reasons, we do not normalize the boundary length with the temperature of the theory in the figures. In Appendix A, we briefly discuss the normalized lengths for completeness.

\begin{figure}[t]
\begin{minipage}[ht]{0.5\textwidth}
\begin{flushleft}
\centerline{\includegraphics[width=80mm]{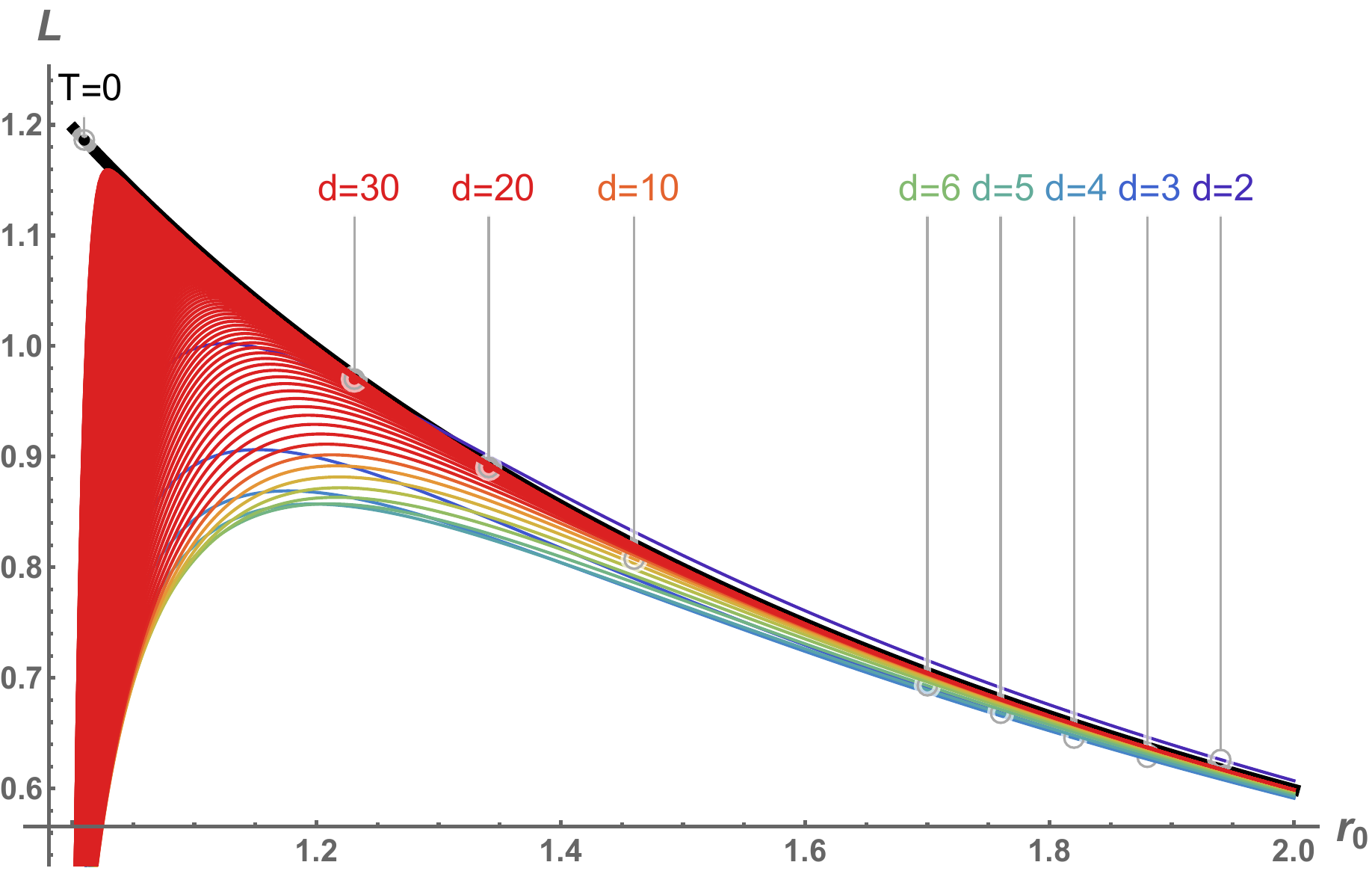}}
\caption{\small{The boundary length dependence on the turning point for surfaces of various dimension. Surfaces that are away from the horizon of the black hole, that is for large $r_0$, behave as being at the vacuum solution even for intermediate values of $d$. As we increase the number of dimensions, the right branch of the curve, which corresponds to the stable minimal surfaces, approaches quickly the $T=0$ surfaces. The computational advantage of the large $d$ expansion lies in the weak dependence of the results on the dimensionality and the preservation of both stable and unstable branches for any $d$. In the plot we present $L(r_0)$ for $d=2$ to $d=250$. Notice the qualitatively different behavior of $d=2,3,4$ curves that appear overlapping the red lines. }}
\label{figure:Lr0t}
\end{flushleft}
\end{minipage}
\hspace{0.3cm}
\begin{minipage}[ht]{0.5\textwidth}
\begin{flushleft}
\centerline{\includegraphics[width=77mm]{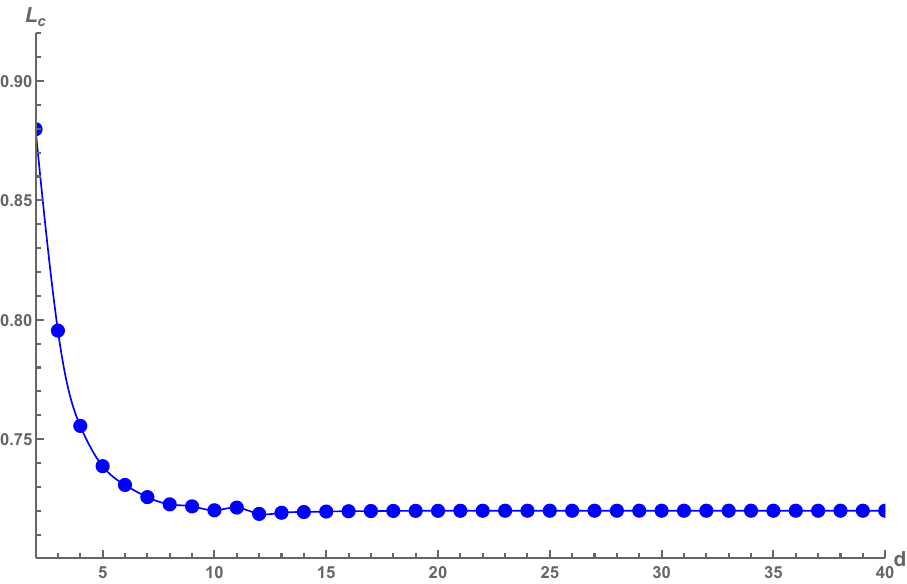}}
\caption{\small{The critical $L_c$, where the meson phase transition happens, is where the connected minimal surface prefers energetically to becomes disconnected with the same boundary conditions. Notice the fast convergence to a plateau value and that a 4-dimensional field theory ($d=4$) is approximated very well by the large $d$ expansion. The form of the curve for large and intermediate $d$ implies that the product $L_c T$ is linear with respect to $d$, confirming the analytic large $d$ analysis \eq{erl2}.\vspace{2.2cm}}}
\label{figure:Lc1}
\end{flushleft}
\end{minipage}
\end{figure}

Alternatively, the described extremal surface dependence can be thought of in terms of fixed boundary length $L$.
A property of the minimal surfaces at finite temperature is the existence of two solutions with different turning point for fixed distance $L$. As we increase $d$, we confirm that the stable branch of our surfaces approaches the $T=0$ solution, while the stable branch of the surfaces is dominating over the unstable. Nevertheless, the unstable branch is always present irrespective of the dimensionality of the theory and shrinks as we increase $d$. This demonstrates the effect of the localization of the gravitational potential in the near horizon regime as $d$ increases. In figure \ref{figure:Lr0t} we plot the $L(r_0)$ dependence on the dimensions and we compare it to the $T=0$ surface. It is worthy to note in the figure that low $d$ surfaces show a slightly different behavior initially.

The discussion so far has been focused on the way that the minimal surfaces behave as we increase the number of dimensions. From the physical point of view, the interest is mainly on the way that the energy of the surface changes with respect to its boundary distance. In other words how the energy of a heavy meson changes as we place the bound state in a space with more dimensions. The main interest is on the behavior of the critical temperature (or the critical length) where the meson melts as we increase the dimensionality of space. That is the critical distance $L_c$, where the energy of the connected minimal surface becomes energetically unfavorable compared to the two disconnected ones with the same boundary conditions. This is the phase transition for the meson.

To analyze this behavior we have to compute numerically the integral of the energy \eq{stot} with respect to the integral of the length \eq{lwl}. In practice we compute the integrals for different turning points of the surface $r_0$ and then we trade $r_0$ with $L$. The critical length decreases with $d$ monotonically, with a quick saturation to a plateau. The saturation point is already achieved for a ten-dimensional field theory. The results appear in Figure \ref{figure:rd1}. To an extent, this behavior reflects the fact that an increasing $d$ can be seen as naively as corresponding to increase of the temperature. The plateau of $L_c$ indicates that the gravitational potential is extremely localized in the near horizon regime where the main contributions to the potential come from the $S_m$ term and in particular of its horizon contribution. This is a main difference with the low temperature expansion. All this is by normalizing the dimensionful quantities with the fixed $r_h$.

The numerical results confirm our analytical expectations \eq{erl} and \eq{erl2}  for large $d$ and extend the analysis to intermediate and low values of $d$. The large $d$ lower bound on $L_c$ is already saturated for $d\simeq 9$. Comparing the critical length $L_c$ of a 1-dimensional field theory to the large $d$ result, we find a modification of about $18\%$, while the 4-dimensional field theory result gets  modified only by $4\%$ in the large $d$ expansion. The applicability of the large $d$ expansion for this observable is tremendous.

An almost perfect fitting can be made for large $d$ on the critical curve \eq{figure:Lc1} as $L_c T \simeq 0.057 d -2 \cdot 10^{-5}  \cO(1/d)$. The leading  coefficient matches the one in the analytic expression \eq{erl2} and the extrapolation to lower values works very well. For example, for $r_h=1$ the extrapolation to $d=4$ produces $L_c=0.229$ compared to the explicit numerical value of $0.240$. The large $d$ expansion can be applied reliably to lower dimensions by extrapolation.

\subsection{Wilson Surfaces of Arbitrary Codimension and the Large $d$ limit}

Let us generalize the analysis to Wilson surfaces of space-time codimension $d-q-1$ (from the boundary point of view). These could be associated with certain configurations of extended objects such as $q$-dimensional branes and anti-branes in the higher dimensional boundary theories. The relevant boundary surface $S$ extends along $q$ spatial directions, the time direction and it is separated along another spatial direction. For $q=0$, the analysis reduces to the Wilson loop. The Wilson surface observables can be computed by minimizing a $(q+1)$-dimensional brane action in the bulk.
The action in static gauge reads
\be
2 \pi {\ell^2} S_A= 2 \cT\, \int g_{xx}^\ff{q}{2} \sqrt{-g_{tt} \prt{g_{xx} + g_{rr} r'^2}}d\s~,
\ee
where $\ell^2 = 1/(T_{q+1}V_q)$ is given in terms of the tension $T_{q+1}$ of the $(q+1)$-brane and the volume $V_q$ of the $q$ spatial directions. 
In the limit $\cT \to \infty$, the quantity $S_A/\cT$ is expected to give the energy of the ''brane-antibrane`` configuration in the boundary theory. As before, we drop an overall factor $\cT$ in the rest of this section.

The first order equation of motion is
\be
r'^2=-\ff{g_{xx}\prt{g_{tt} g_{xx}^{q+1}+c^2}}{c^2 g_{rr}}~,
\ee
where in terms of the turning point $r_0$ we define  $c^2=-g_{tt}(r_0) g_{xx}^{q+1}(r_0)$. For practical reasons the integration can be expressed with respect to $r$ as
\be\la{sq}
2 \pi \ell^2 S_A=2\int g_{xx}^q \sqrt{-\ff{g_{tt}^2g_{rr} g_{xx}}{g_{tt} g_{xx}^{q+1}+c^2} } dr~.
\ee
As before we will consider the quantity $S_{tot}=S_A-S_m$, where  
\be
2 \pi \ell^2 S_{m}=2 \int \sqrt{-g_{tt} g_{xx}^q g_{rr}} dr~,
\ee
which is free of any UV divergences as in eq \eq{sa12}.   The separation length on the boundary for the surface is
\be
L=2 \int \sqrt{\ff{c^2 g_{rr}}{-g_{xx}\prt{g_{tt} g_{xx}^{q+1}+c^2}}} dr~.~
\ee
The application of the generic formalism on the gravity dual background gives
\be
L=2\int_0^1 \ff{y^{q+2}\sqrt{1-w^d}}{r_0\prt{1-w^d y^d}}\ff{1}{\sqrt{1-\ff{\prt{1-w^d} y^{2\prt{q+2}}}{1-w^dy^d}}} dy
\ee
and for the action
\be
2 \pi \ell^2 S_A=2\int_0^1\ff{r_0^{q+1}}{y^{q+2}\sqrt{1-\ff{y^{2\prt{q+2}}\prt{1-w^d}}{1-w^d y^d}}} dy~,
\ee
where $w:=r_h/r_0$ and $y:=r_0/r$, both of which are less or equal to one. The above expressions are written in such a way that we can consecutively use the binomial theorem to get
\be\la{al1}
L=\ff{2}{r_0}\ff{1}{d m+1+\prt{2 +q}\prt{2 n+1}} c_n b_{m,n} w^{d m}\prt{1-w^d}^{n+\ff{1}{2}}~,
\ee
where
\bea
c_n:=\binom{n-\ff{1}{2}}{n}=\ff{\Gamma\prt{n+\ff{1}{2}}}{\sqrt{\pi} \Gamma{\prt{n+1}}}~,\qquad b_{m,n}:=\binom{m+n}{m}~,
\eea
while repeated indices are summed from zero to infinity. The minimal action now is equal to
\be
2 \pi \ell^2 S_A=\ff{2 r_0^{q+1}}{ d m+1+\prt{2 n-1} \prt{q+2}} c_n b_{m,n-1} \prt{1-w^d}^n w^{d m}\prt{1-y^{d m+1+\prt{2 n-1} \prt{q+2}}\big|_{y=0}}~.
\ee
The divergence in the expansion has been isolated in the second term in the brackets for $m=n=0$, for any value of $q$. This term is the one that gives the infinity and cancels against the renormalization term, while for $m,n\neq 0$ it is null. Therefore, $S_{tot}$ can be written as
\be\la{ad1}
2 \pi \ell^2 S_{tot}=  \ff{2 r_0^{q+1}}{ d m+1+\prt{2 n-1} \prt{q+2}} c_n b_{m,n-1} \prt{1-w^d}^n w^{d m}+\ff{2r_h^{q+1}}{q+1}~.
\ee
Notice the elegance of this exact finite expression when written in this form. We highlight that the finiteness of this expansion relies on the fact that the divergences turn out to be isolated in single terms as we have shown. Moreover, in the limit $q\rightarrow 0$ is smooth and corresponds to Wilson loop.

In the large $d$ limit the main contribution comes from $m=0$, and we can simplify significantly both expressions \eq{al1} and \eq{ad1} by reducing the number of independent summations as
\be\la{al12}
L\simeq\ff{2}{r_0}\ff{1}{1+\prt{2 +q}\prt{2 n+1}} c_n \prt{1-w^d}^{n+\ff{1}{2}}
\ee
and
\be
2 \pi \ell^2 S_{tot}\simeq\ff{2 r_0^{q+1}}{1+\prt{2 n-1} \prt{q+2}} c_n \prt{1-w^d}^n +\ff{2r_h^{q+1}}{q+1}~.
\ee
For completeness, one may confirm the validity of the above expressions by making an even more drastic approximation at large $d$, as $(1-w^d)^n\simeq 1$, to obtain
\be
r_0\simeq \ff{2 c_n}{\prt{3+q+ 2 k(2+k) }L}
\ee
and
\be
\ell^2 S_{tot}\simeq\ff{2^{1+q}c_n c_k^{1+q}}{\prt{3+q+2 k(2+q)}^{1+q}\prt{1+q-2 n(2+q)}\pi L^{1+q}}+\ff{r_h^{q+1}}{\pi\prt{q+1}}~.
\ee
Notice that the last two expressions, obtained by the drastic approximation, serve as a validation of the approximation and can be obtained analytically in other ways.

For $q=0$ the expansion is valid for the Wilson loops, where already the first term of the summation, $k=n=0$, provides accuracy up to the first decimal digit. Roughly speaking, by increasing the order of magnitude of the summed terms by one, the accuracy is improved by one decimal digit, compared with the expression \eq{stota1}.

We have studied the properties of the Wilson loop for a large $d$ dual field theory in the deconfined phase. The Wilson loop in the confining phase has a different behavior, which originates from the properties of the minimal surfaces in geometries which exhibit confinement.

\section{Confining QCD Strings at Large $d$}

The fluctuations of the flux tube of a meson in conformal theory is independent of the number of dimensions. Therefore at zero temperature and in the absence of any scale in the theory, the Wilson loop should remain unaffected by the number of dimensions. If a scale is present, a temperature or a mass gap, the properties depend on it and also on the dimensionality of the theory. In this section we compute the holographic Wilson line for a confining theory at large $d$.

The soliton solutions \eq{soliton} exhibit confinement at long distances and Coulombic behavior at short distances, a qualitative behavior that matches the QCD lattice calculations. Here we discuss whether the large $d$ behavior preserves these features, and while preserving them, if there is a mechanism capable of extracting these features in a manner simpler than the finite $d$ holographic computation. 

The analytical manipulation of \eq{sa1} and \eq{sren} leads to the following expression for the potential
\be\la{confi}
2\pi \a' S_{tot}= c L+2\prtt{\int_{r_0}^\infty \sqrt{-g_{rr} g_{tt}}\prt{\sqrt{1+\ff{c^2}{g_{xx} g_{tt}}}-1}-\int_{r_k}^{r_0} \sqrt{-g_{00} g_{rr}}}:= c L+ K(r_0)~,
\ee
where c is given by \eq{equ} and it is proportional to the string tension. The discussion boils down to the behavior of the terms in the brackets as the dimensionality increases. The linear meson potential is recovered for large $L$, where $r_0\simeq r_k$.  The terms in the brackets for the surfaces sitting at the tip of the geometry give a constant finite contribution $K(r_k)$. Once we increase the number of dimensions, this remains finite and converges to a constant value. The string tension reads
\be\la{ts1}
\s=\ff{c}{2 \pi \a'}  =\ff{r_k^2}{2 \pi \a'}~
\ee
and it is independent of the dimensionality of the theory.

\begin{figure}
\begin{minipage}[ht]{0.5\textwidth}
\begin{flushleft}
\centerline{\includegraphics[width=75mm]{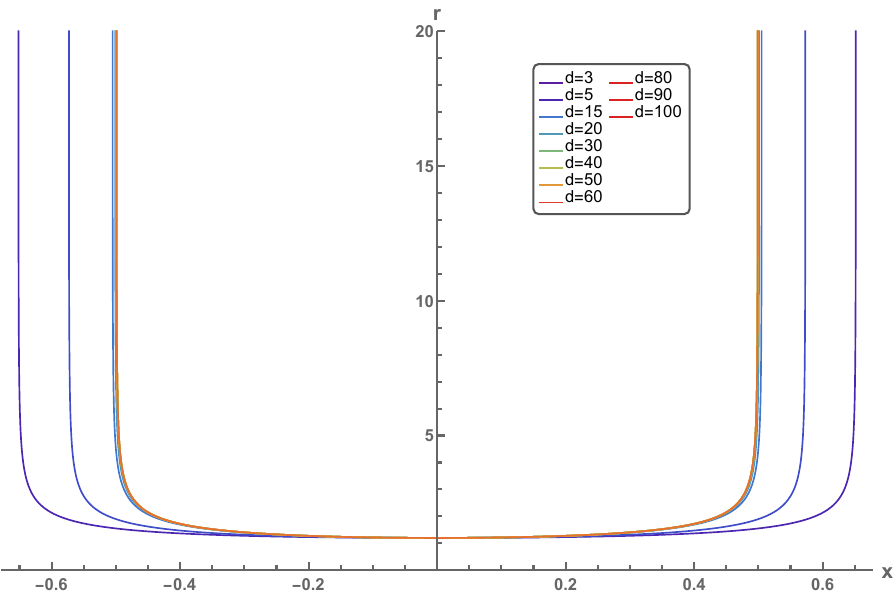}}
\caption{\small{The shape of the surfaces with fixed turning point $r_0=1.2 r_k$.
As we increase the dimensionality of the theory, the minimal surface converges to a minimal surface and already for $d\ge15$, we observe no change on the surface.\vspace{.3cm}}}
\label{figure:cons1}
\end{flushleft}
\end{minipage}
\hspace{0.3cm}
\begin{minipage}[ht]{0.5\textwidth}
\begin{flushleft}
\centerline{\includegraphics[width=75mm ]{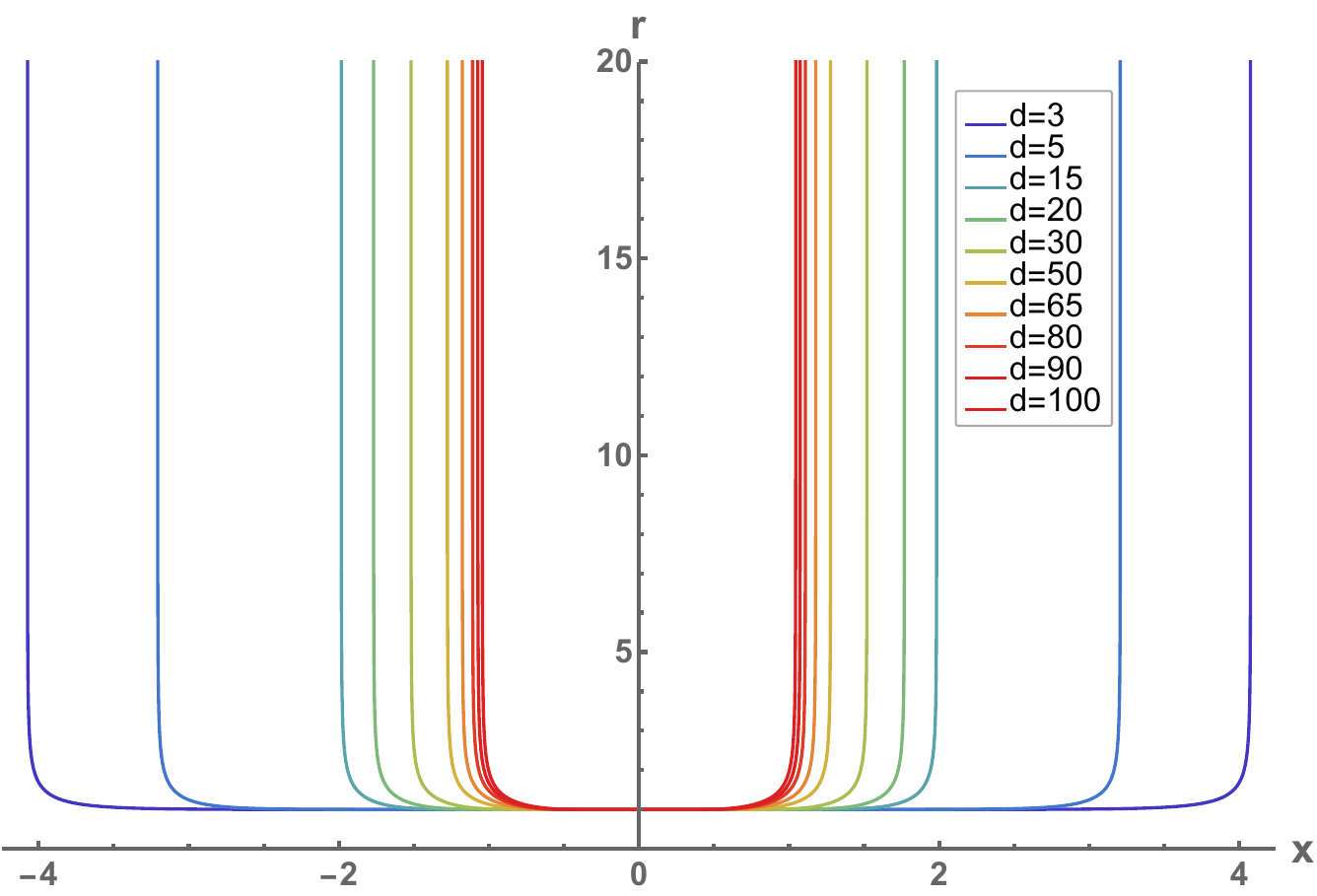}}
\caption{\small{The surfaces with turning point at the tip of the theory, in the plot for $r\simeq \prt{1+ 10^{-6}}r_k$. This is the limit of large $L$. The closest we approach the $r_k$, the larger the boundary length gets. For larger dimensions, we need to approach closer to the tip to achieve the large interquark distances.}}
\label{figure:cons2} \vspace{.0cm}
\end{flushleft}
\end{minipage}
\end{figure}

Let us briefly study numerically the geometric properties of the confining minimal surfaces at large $d$. Here we have $d\ge3$ since one of the spatial dimensions $\phi$ has been compactified. For a theory of fixed dimension, we compute numerically the Wilson loop for a wide range of interquark distances in order to probe the regime of the linear potential and the regime of the Coulombic behavior. The Wilson line surface has a unique solution in the solitonic geometry. We solve the differential equation \eq{equ} in the Coulombic regime at small distances and we present its shape in figure \ref{figure:cons1}. The minimal surface at large distances, $r\simeq r_k$, is presented in figure \ref{figure:cons2}, where it has the characteristic behavior of an inverse $\Pi$ lying to large extend on the cut-off scale $r_k$ of the theory. This behavior is universal as long as the dual theory exhibits confinement and it is independent of the dimensionality of the theory. It is the geometric feature of Wilson loop that generates confining behavior. The two straight lines of the inverse $\Pi$ are subtracted by the infinite masses of the quarks and one remains with a ``flat" string at $r=r_k$ which leads to confinement.  In summary, we observe that the shape of the surface related to the confining Wilson loop is approximately independent of the dimensionality of the theory for large $d$, while as $d$ increases the surfaces converge to a certain surface.

\begin{figure}
\centerline{\includegraphics[width=82mm]{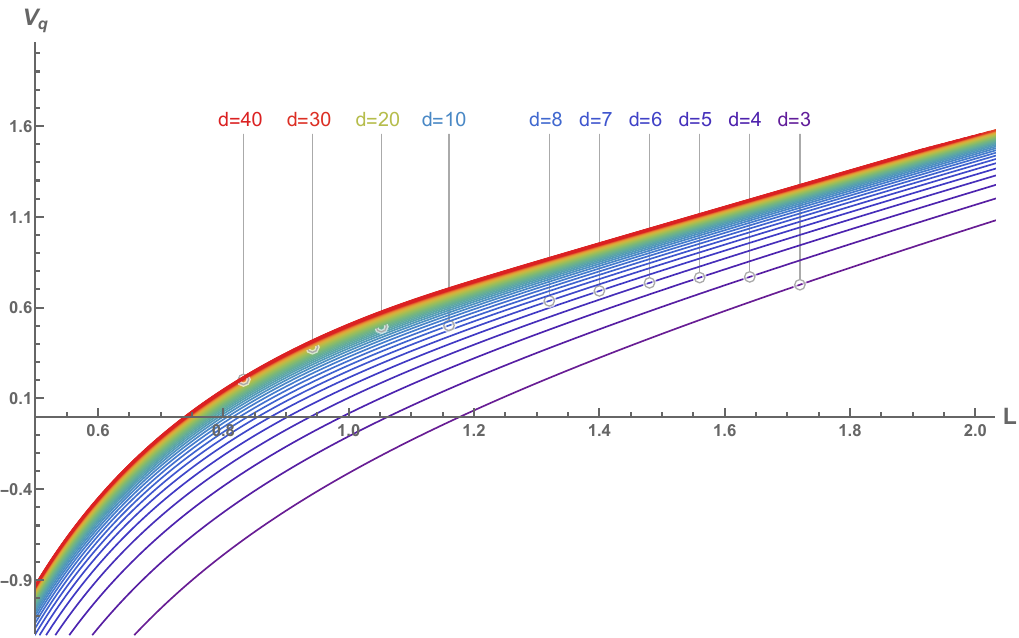}}
\caption{\small{The quark potential for different dimensions. At smaller distances $L$, the potential is of Coulomb type, while for larger ones it becomes linear, revealing the confining nature of the theory. We note the quick convergence to a certain stable form of the potential plotted here in red. The convergence is achieved already for low dimensionality, as of around $d=11$, where beyond that minimal changes are observed. This is demonstrated in the plot by the density of $V_q$-lines in the intermediate and large $d$ regime. Notice that the gradient of the straight line for large $L$, which corresponds to the string tension, remains unchanged for different dimensionality. The lower dimensions are the ones with the lower potential in the plot and the rainbow color convention is followed as we increase the dimension of the theory.}}
\label{figure:vqc}
\end{figure}

We numerically compute the interquark potential with respect to its distance $L$ at various dimensions.  The linear confining potential for large $L$ and the Coulombic potential for small interquark distances is found irrespectively of the dimensionality of the theory, as expected. The numerical evaluation is shown in figure \ref{figure:vqc}. The interquark potential converges at large $d$ to a given potential. The convergence happens quickly and already for $d\sim 14$, we have reached a regime where further changes are not noticeable as we increase the dimensionality. The slope of the potential determining the string tension $\s$ is constant, and matches \eq{ts1}, while the constant contribution $K(r_k)$ has converged to a given value already.

The large $d$ behavior determines with good accuracy the string tension $\s$ while the full potential $V$ is only slightly modified compared to that of the lower dimensions. In particular, for a 4-dimensional confining field theory, we find that in the Coulombic regime the physical constant coefficient of the $L^{-1}$ term in the potential differs only by about $3\%$, while the constant term differs by about $30\%$ compared to the large $d$ expansion. In the large $L$ regime the string tension is $d$-independent, while the constant $L$-independent term differs by $40\%$.

In summary, we see a quick convergence of our physical quantities in the large $d$ regime. The heavy quark physics in the large $d$ regime continues to capture all the qualitative features of the low $d$ regime, including confinement and the Coulombic behavior. This is of course not unexpected. Moreover, we observe that the physically interesting quantities depend weakly on the dimensionality of the theory. Our analysis strongly suggests that a wider application of the large $d$ expansion even in solitonic backgrounds, not only in black holes, can be useful.

\section{Entanglement Entropy at Large $d$}

\subsection{Entanglement Entropy at Large $d$ Analytically}

In this section we study a different type of extremal surfaces, dual to the entanglement entropy. Let us apply the coordinate transformation $z=1/r$ to bring the boundary of \eq{met11} at $z=0$. To keep track of dimensions of various quantities, let us restore the radius of curvature of $AdS$ space. The metric then takes the form
\be
ds^2_{d+1}=\ff{R^2}{z^2} \prt{-f(z) dt^2+dx_{d-1}^2+\ff{dz^2}{f(z)}}~, \qquad f(z)=1-\ff{z^d}{z_h^d}~.
\ee
We consider a strip sub-region for the entanglement entropy and follow the usual minimization procedure for the area. The strip is the boundary of the entangling surface comprising two parallel planes of spatial co-dimension one, separated by a distance $L$ along the $x$-direction. The minimization of the area reads
\be\la{ees}
4 G_N^{(d+1)} S=\cR^{d-2} \int d\s g_{xx}^{\ff{d-2}{2}}\sqrt{g_{xx}+g_{zz}z'(\s)^2}~,
\ee
where $\cR^{d-2}$ is the $(d-2)$-dimensional spatial volume transverse to the $x$ direction and $G_N^{(d+1)}$ is the Newton's constant in $d+1$ dimensions. The length $L$ of the strip  extending along the $x$-direction is obtained by the manipulation of $z'(\s)$ in the first order equation of motion
\be
z'(\s)^2= \ff{g_{xx}\prt{g_{xx}^{d-1}-c^2}}{c^2 g_{zz}}~,
\ee
where the constant $c^2:=g_{xx}^{d-1}(z_0)$ with $z_0$ being the turning point of the surface. The length $L$ is given by
\be
L=2 \int  \sqrt{\ff{g_{zz}}{g_{xx}^{d} g_{xx}(z_0)^{1-d}\prt{1-g_{xx}^{1-d} g_{xx}(z_0)^{d-1}}}} dz~.
\ee
The minimization integrals for the background under study become
\be\la{lee}
L=2 \int_0^{z_0} \ff{z^{d-1}}{z_0^{d-1}}\ff{1}{\sqrt{f(z) \prt{1- \prt{\ff{z}{z_0}}^{2\prt{d-1}}}}}~.
\ee
The extremal area reads
\be\la{seeh}
4 G_N^{(d+1)} S=2 R^{d-1} \cR^{d-2}\int_{\e}^{z_0} dz \ff{1}{z^{d-1} \sqrt{f(z)\prt{1-\prt{\ff{z}{z_0}}^{2\prt{d-1}}}}}~.
\ee
The divergence of the integrand is of the order of $z^{1-d}$. Therefore, we add and subtract the relevant term to isolate the divergence, integrate the extra term from $\e/z_0$ to 1, and obtain the known result
\be\la{sentropy}
4 G_N^{(d+1)} S= R^{d-1} \cR^{d-2}\prtt{\ff{2}{\prt{d-2}\e^{d-2}} +\ff{2}{z_0^{d-2}}\prt{-\ff{1}{d-2}+\int_{0}^{1} dy \ff{1}{y^{d-1}}\prt{\ff{1}{ \sqrt{f(y z_0)\prt{1-y^{2\prt{d-1}}}}}}-1}}~.
\ee
To extract the large $d$ contributions, we write the area in the following form
\be\la{eearea00}
S= \ff{R^{d-1} \cR^{d-2}}{4 G_N^{(d+1)}}\prtt{ \ff{2}{\prt{d-2} \e^{d-2}}+ \ff{L}{z_0^{d-1}}+\ff{2}{z_0^{d-2}}\prt{-\ff{1}{d-2}+ F(z_0)}}~,
\ee
where
\be\la{fzoo}
F(z_0)=  \int_0^{1} dy \ff{1}{y^{d-1}}\prt{\sqrt{\ff{1-y^{2\prt{d-1}}}{f(z_0 y)}}-1}
\ee
is the finite term that depends on the turning point $z_0$, and therefore, the length $L$ of the surface. We expect the factor $R^{d-1}/G_N^{(d+1)}$ to be proportional to a power of the number of degrees of freedom of the boundary theory. The advantage of the extremal area written in the form \eq{eearea00} is that we have isolated the divergence in the first term, and we have extracted the dependence on $L$, $z_0$ and $d$ of the rest of the terms.

For large separations $L$, corresponding to surfaces that probe the IR, $z_0\simeq z_h$, the finite part of the area takes the form
\be\la{szh}
\ff{4 G_N^{(d+1)} S_{finite}}{R^{d-1}\cR^{d-2}} \simeq \ff{L}{ z_h^{d-1}}  + \frac{2}{z_h^{d-2}} \prt{-\ff{1}{d-2}+F(z_h)}~.
\ee
The first term is linear to $L$ with a proportionality factor of $z_h^{1-d}$.  Let us consider the large $d$ limit of the $L$ independent term. The factor $-\ff{1}{d-2}$ is of order $\cO(d^{-1})$, while $F(z_h)$ reads
\be\la{fzh}
F(z_h)=  \int_0^{1} dy \ff{1}{y^{d-1}}\prt{\sqrt{\ff{1-y^{2\prt{d-1}}}{1-y^d }}-1}~.
\ee
By factorizing the square root and taking the large $d$ limit in the square 
the integral can be done analytically and expressed in closed form in terms of hypergeometric functions
\be
F(z_h)\simeq \ff{y^{-d}}{d}\prt{1-{}_2 F_1\prtt{-\ff{1}{2},-1,\ff{2}{d},-y^d}}\bigg|_{y=0}^{y=1}~,
\ee
which can be seen that for large dimensions it just  converges to a $d$ independent value as, giving for $F(z_h)$
\be\la{fres}
F(z_h)\simeq \ff{1}{4}~.
\ee
This result is essential for the area theorem which we discuss later in the section and  in the section with the numerical analysis.

At this point it is instructive to define the relative measure of entanglement in an excited state compared to the vacuum state of the CFT. An appropriate definition \cite{Gushterov:2017vnr} is
\be\la{sdef}
S_{density}=\ff{S-S_0}{V}~,
\ee
where $S$ and $S_0$ are the corresponding entanglement entropies. $S_{density}$ is finite and cut-off independent.  The short distance divergences are identical in both theories, since the corresponding UV fixed points are the same. Using \eq{szh} and the vacuum state entanglement coming from the integration of \eq{seeh}, one may obtain $S_{density}$ for large $L$ and large $d$ to be equal to
\be\la{sde}
4 G_N^{(d+1)} S_{density}\simeq R^{d-1}\prtt{\ff{1}{z_h^{d-1}}+\ff{2}{z_h^{d-2} L}\prt{-\ff{1}{d-2}+F(z_h)}}~,
\ee
where we have utilized properties of the $\Gamma$ functions \eq{gamma1} and that $Vol=\cR^{d-2} L$.  
In the large $L$ limit, the entanglement entropy $S$ approaches the thermal entropy $s \,Vol=R^{d-1}Vol/4 G_N^{(d+1)} z_h^{d-1}$, since the entangling surface covers the whole space, with subleading corrections. In this limit it takes the form
\be \la{eeee}
S  \simeq s Vol +\a Area~,
\ee
where the subleading second term contains a dimensionful constant $\a$ which is known to obey a monotonicity theorem along the RG flow, known as the area theorem \cite{Casini:2012ei,Casini:2016udt}.

The validity of the area theorem can be understood as follows. In the CFT vacuum there are no scales beyond the CFT cut-off and the length of the strip, so the form of the entanglement entropy is determined by dimensional grounds. In the excited CFTs, there is a UV contribution that matches to the CFT vacuum one since the theories have the same UV fixed point. Therefore, the divergent contributions in $\a$ included in the numerator of \eq{sdef} cancel against each other.  There is a variety of additional finite terms allowed by the scales in the theory that contribute to $\a$. In the presence of Lorentz symmetry, it is known that the coefficient of these terms follows a holographic $c$-theorem \cite{Ryu:2006ef,Myers:2010tj,Chu:2019uoh,Giataganas:2017koz,Cremonini:2013ipa}. In the entropy density \eq{eeee} defined above, the contributions of $\a$ can be understood as the difference $\a_{IR} - \a_{UV}$, and therefore, whenever the area theorem holds, the difference has to be negative.

From \eq{eeee}, we get
\be
\ff{S_{density}}{s}\simeq 1+ \prt{\a_{IR} - \a_{UV}}\ff{4 G_N^{(d+1)} z_h^{d-1} A}{R^{d-1}}~,
\ee
similar to \cite{Gushterov:2017vnr}, which upon a substitution to \eq{sde}, and at large $d$, leads to a simple dependence on the integral $F(z_h)$ as $\a_{UV}=\a_{IR} - s z_h F(z_h)~$.
At large $d$ the area theorem violation boils down to the computation of the sign of $F(z_h)$. The large $d$ expansion \eq{fres} shows that $F(z_h)$ converges to a positive value, and therefore, at large $d$ the area theorem can be seen analytically that is violated
\be
\a_{UV}=\a_{IR} -\ff{s z_h}{4}~, \qquad\mbox{at large $d$}~.
\ee
The robustness of the large $d$ expansion is tied with the straightforward analytic proof of the
area theorem violation. There is no need for numerical analysis or complicated analytic computations. The fact that we can analytically compute the function $F(z_h)$ in this limit and conclude for the area theorem in a relatively straightforward way demonstrates one of the benefits of the large $d$ expansion, and its possible applicability to other theories and other monotonicity theorems. In subsection \ref{section:qen} we will elaborate more on the area term computation and we will provide an alternative evaluation of the $\a$ values.

Before we conclude this section, let us  also briefly comment on the properties of the entangling surface. The area written as \eq{eearea00} depends on $L$ through $z_0$; that is, through an expression that is not of a closed form. The entangling surface at large $d$ is given by the Euler-Lagrange equation \eq{ees}
\be\la{zee}
z'^2(\s)= f\prt{z(\s)} \prt{\ff{z_0^{2\prt{d-1}}}{z^{2\prt{d-1}}}-1}~.
\ee
The differential equation has two opposite sign solutions for the derivative $z'(\s)$, corresponding to symmetric branches of the string solution with respect to the point $x=\s=0$. The derivative of $z$ diverges when the surface becomes transverse to the boundary for $x=\pm L/2$, where $L$ is given by  \eq{lee}. At large $d$, even for intermediate distances from the boundary, one gets a temperature-independent approximation as $z(\s)\sim z_0 d^{1/d} \prt{\s/z_0+c_1}^{1/d}$~, where $c_1$ is the integration constant. The constant may be determined by symmetry, as $c_1\simeq 1/d$, since $z(0)=z_0$. The boundary length $L$ given by the solution $z(L/2)=0$ is found to be inversely proportional to the dimension as
\be\la{lzd}
\ff{L}{2} \sim \ff{z_0}{d}~,
\ee
for large $d$.  Equation \eq{lzd} is not unexpected, and agrees with the entanglement entropy for the straight strip at zero temperature. There the integral \eq{lee} is analytically doable \cite{Ryu:2006ef}
\be\la{lrule1}
 \ff{L_{T=0}}{2}= \ff{\G\prt{\ff{d+1}{2d}}}{\G\prt{\ff{1}{2d}}}~.
\ee
The large $d$ limit of this expression for surfaces with any turning point, since there is no horizon scale in the theory, gives
\be\la{lrule2}
\ff{L_{T=0}}{2}\simeq \ff{\pi}{2} \ff{z_0}{d}~,
\ee
where we have used the expansions of the $\Gamma$ functions:
\bea\la{gamma1}
&&\G\prt{\ff{d+1}{2d}}=\sqrt{\pi}+\ff{\sqrt{\pi}}{2 d}\ff{\pp}{\pp z}\log \G(z)\bigg|_{z=1/2}+\cO(d^{-2})~,\\
&&\G\prt{\ff{1}{2d}}=-\gamma +2 d +\ff{6 \g^2 +\pi^2}{24 d}+\cO(d^{-2})~,
\eea
where $\gamma$ is the Euler's constant.  Therefore, for large $d$, the expectation is that the extremal surfaces follow an inverse dimension rule which we study below numerically for the whole range of values of $d$.

\subsection{Numerical Analysis at Low, Intermediate and Large $d$}

With the entanglement entropy computed analytically at large $d$, and certain $L$ limits,  we now compute the integrals for the whole range of dimensions. The intermediate and low $d$ regimes are not analytically tractable, while at the large $d$ regime the numerics confirm the analytic results.

\begin{figure}
\begin{minipage}[ht]{0.5\textwidth}
\begin{flushleft}
\centerline{\includegraphics[width=75mm]{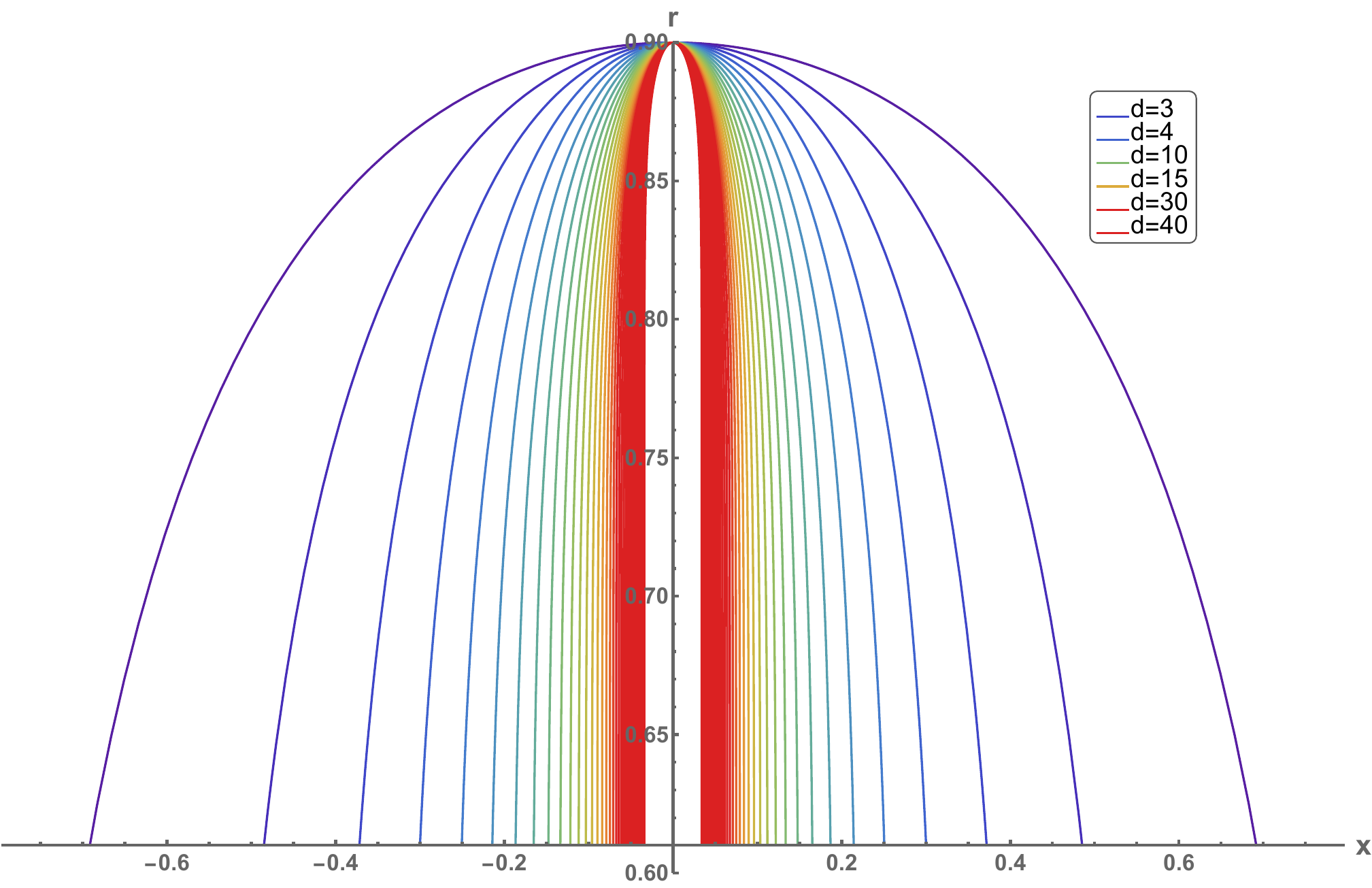}}
\caption{\small{The shape of the entangling surfaces with fixed turning point $z_0=0.9 z_h$ for $d=3$ to $d=40$. The legend contains only few representative values of all dimensions plotted, to make the color correspondence obvious. As the dimensionality increases, the extremal surface requires smaller boundary distance to reach a fixed radial location in the bulk. At higher dimensions, depicted with red, we note an increased density of surfaces, reflecting their weaker dependence on the number of dimensions, due to the localization of the gravitational potential in the near horizon regime.}}
\label{figure:eell}
\end{flushleft}
\end{minipage}
\hspace{0.3cm}
\begin{minipage}[ht]{0.5\textwidth}
\begin{flushleft}
\centerline{\includegraphics[width=75mm ]{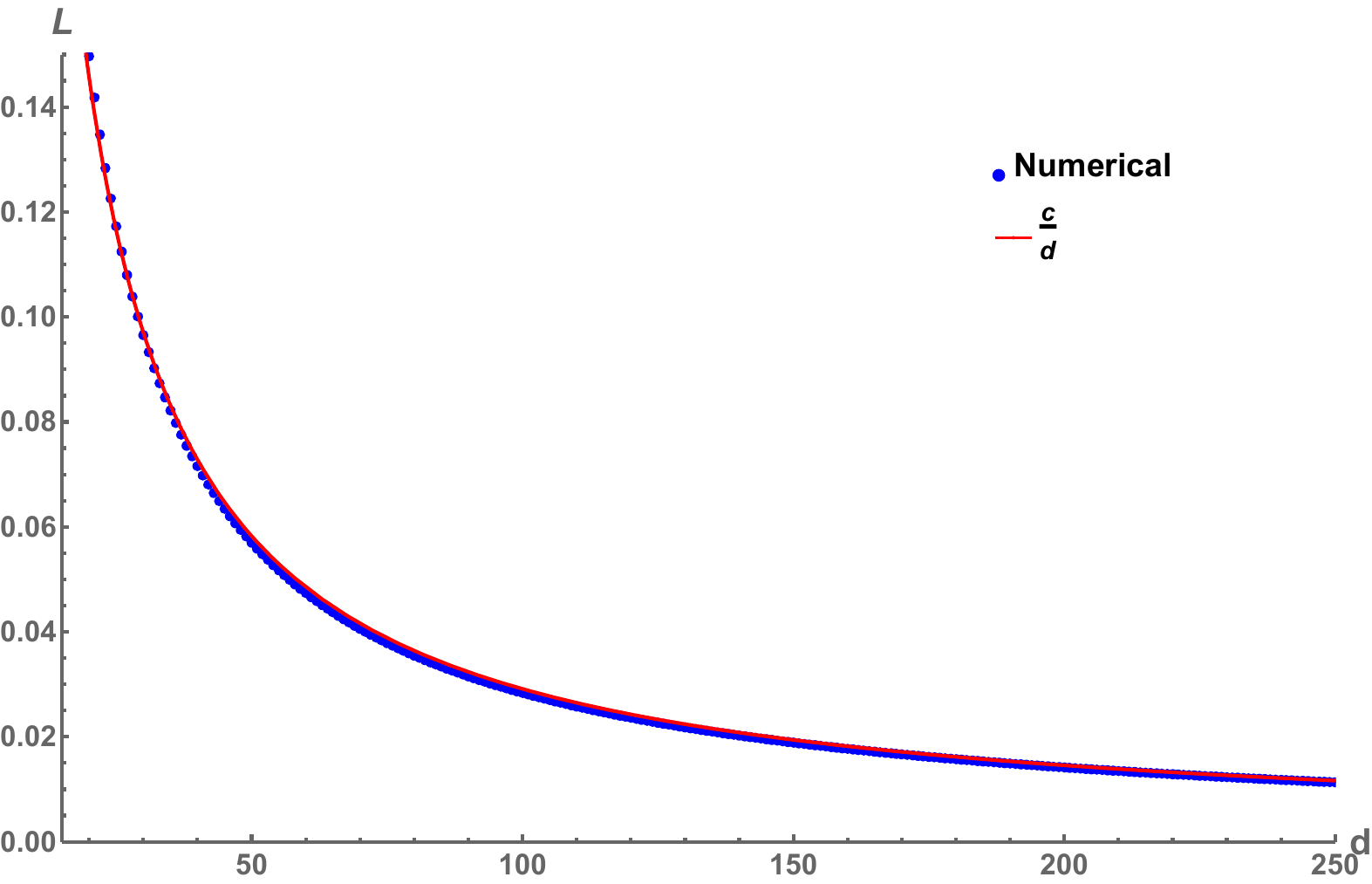}}
\caption{\small{The length $L$ of the extremal surfaces with turning point $z_0=0.9 z_h$  as the dimensionality of the theory increases. For values of $d\sim30$  and above, we observe an exact fitting for $L\sim \ff{1}{d}$, while even for lower dimensions we get approximately a similar behavior with an accurate fitting. In the figure we present the numerical evaluation of $L(d)$ to show its matching with the inverse dimension law $c_1/d$, which is fitted. The plot begins for intermediate values $d=15$ up to the large $d$ limit.\vspace{.9cm}}}
\label{figure:rd22} \vspace{.0cm}
\end{flushleft}
\end{minipage}
\end{figure}

Let us look at the solutions of equation \eq{zee} as we vary the dimensionality. It has two opposite sign solutions for  $z'(\s)$, corresponding to symmetric branches of the surface with respect to $x=\s=0$. We are shooting from the turning point of the surface toward the boundary. The boundary points $x=\pm L/2$, where $L$ is given by  \eq{lee}, are the points where the derivative $z'$ diverges, since the surface becomes transverse to the boundary. We are fixing the tuning point $z_0$ and we are looking at the solution at different dimensions. As we increase the number of dimensions, the extremal surface requires smaller boundary distance to probe the same point in the bulk. For low dimensions we see noticeable changes of the surface as we vary $d$. For higher dimensions the dependence on $d$ becomes weaker. Such a representative set of solutions are plotted in figure \ref{figure:eell}.

The boundary length $L$ decreases with the dimensionality, while the dimensionless quantity $L T$ remains constant.  The inverse $d$ law has been derived for surfaces that have turning points close to the boundary \eq{lrule1} and was further supported by the large $d$ expansion of the analytic vacuum solution. We notice that the function $L(d)$ for intermediate dimensions and surfaces that probe the near horizon regime exhibits similar behavior. In particular, in the large $d$ approximation
\be
L T=\ff{c_1}{4 \pi}+c_2 \ff{d}{4 \pi} +c_3 \ff{d^2}{4 \pi}
\ee
the numerical coefficients $c_2$ and $c_3$ turn out to be of order $10^{-4}$ and $10^{-6}$, practically negligible, while $c_1$ is of order 1, as expected.

The entangling integrals are of higher dimensionality and therefore depend stronger on the dimensionality of the theory compared to the two-dimensional Wilson loop surfaces, as it can be seen from our numerical analysis. However,  we still observe that the large $d$ expansion already captures the qualitative characteristics of the extremal surfaces in the low $d$ regime as shown in figure \ref{figure:rd22}.

\begin{figure}
\begin{minipage}[ht]{0.5\textwidth}
\begin{flushleft}
\centerline{\includegraphics[width=75mm]{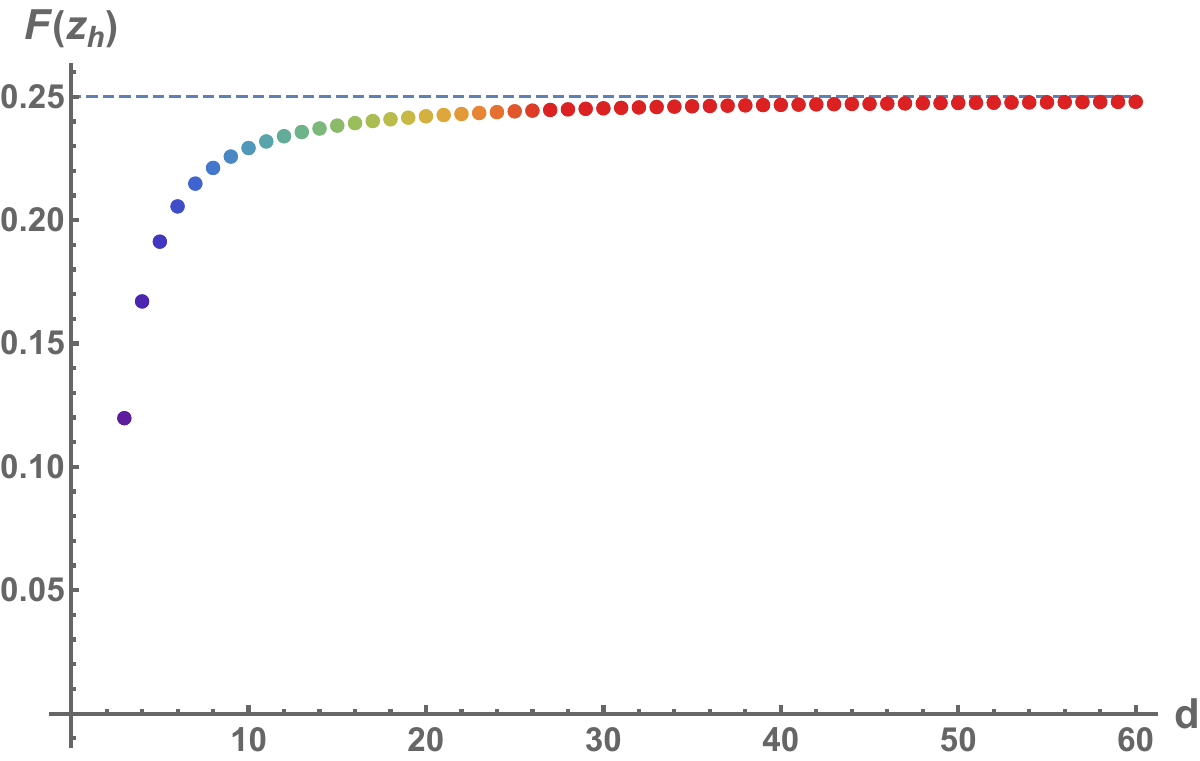}}
\caption{\small{The dependence of the function $F(z_h)$ on the number of dimensions of the theory. The dashed line is the analytical large $d$ limit of \eq{fres}. $F(z_h)$ approaches quickly the plateau value for intermediate dimensionality. The color gradient corresponds to increasing $d$ and demonstrates convergence to $1/4$.\vspace{.7cm}  }}
\label{figure:fzh}
\end{flushleft}
\end{minipage}
\hspace{0.3cm}
\begin{minipage}[ht]{0.5\textwidth}
\begin{flushleft}
\centerline{\includegraphics[width=80mm ]{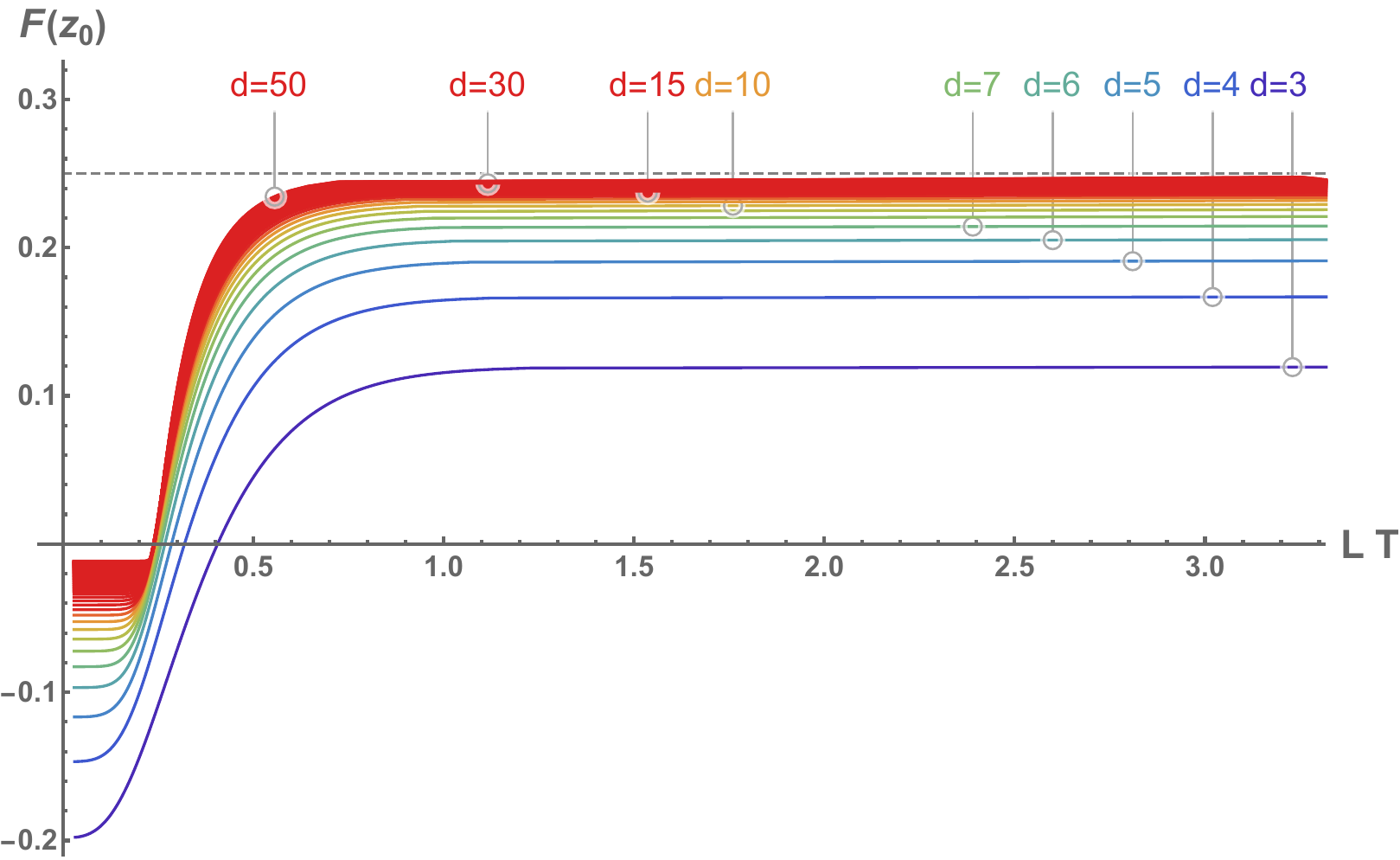}}
\caption{\small{The dependence of the function $F(z_0)$ on the dimensionality of the theory. Every single curve $F(z_0)$ is made for surfaces with the same range of turning points from the near horizon regime to the near boundary regime, while the dimension $d$ is kept fixed. The coloring labels the function $F(z_0)$ at different dimensions. As $d$ increases, $F(z_0)$ is modified at a slower rate as can be seen by the plotted red curves.}}
\label{figure:fz0} \vspace{.0cm}
\end{flushleft}
\end{minipage}
\end{figure}

Let us now turn our discussion to the entanglement entropy. We  look at the simpler case of entangling surfaces that probe the horizon of the black hole. For these type of surfaces, we have shown that the finite part of the entanglement entropy is proportional to $L$ as \eq{szh}
\be
\ff{G_N^{(d+1)} z_h^{d-1}}{R^{d-1} \cR^{d-2}} S_{finite}\simeq L +2 z_h\prt{-\ff{1}{d-2}+ F(z_h)} ~.
\ee
Expressed in the above form the linear behavior in $L$ is $d$-independent, and the dependence on dimensionality comes non-trivially via the $F(z_h)$ function. The analytical expression \eq{fres} of the $F(z_h)$ at large $d$ shows convergence to  $1/4$. The numerical analysis presented in figure \ref{figure:fzh} confirms the analytical expansion for large $d$. In fact the integral approaches quickly its stable $d$ independent value and then it depends weakly on $d$:
\be
\ff{F_{d=4}(z_h)}{F_{d\rightarrow \infty}(z_h)}\simeq 0.67~,\quad \ff{F_{d=5}(z_h)}{F_{d\rightarrow \infty}(z_h)}\simeq 0.86~,\quad \ff{F_{d=15}(z_h)}{F_{d\rightarrow \infty}(z_h)}\simeq 0.95~,
\ee
which performs very well even as a quantitative extrapolation. Moreover, it suggests that the large $d$ expansion is a useful tool for the search and study of monotonicity theorems.

\begin{figure}
\centerline{\includegraphics[width=82mm]{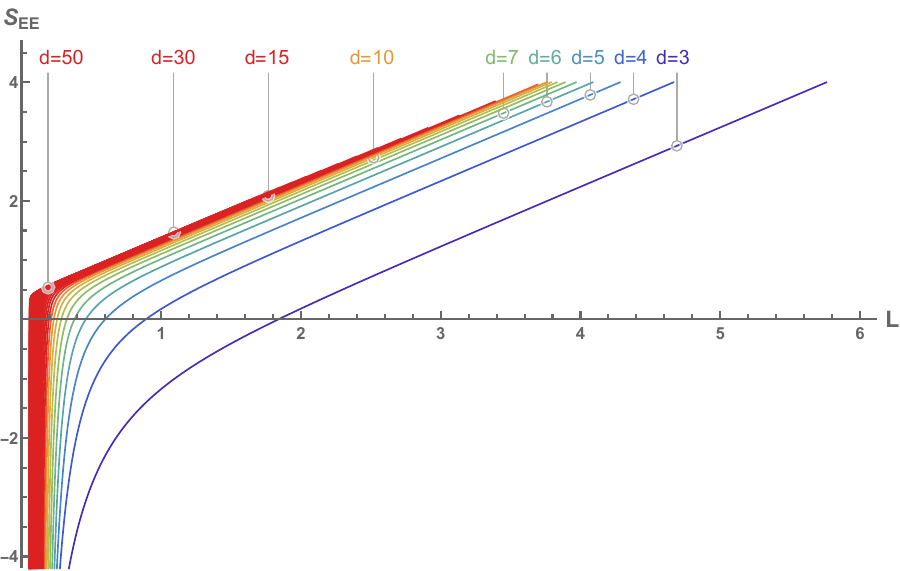}}
\caption{\small{The entanglement entropy for different dimensions. The cut-off has been  subtracted and we keep only the finite part. The entanglement is plotted for all dimensions for a fixed range of the radial coordinates. The increase of $d$ does lead to a certain converging form for the entanglement entropy. In the legend we present some representative values of the dimensions. The plot is done for $d=3$ to $d=50$.}}
\label{figure:eLr1}
\end{figure}

We also solve \eq{fzoo}, for entangling integrals \eq{sentropy} that probe away of the horizon of the black hole for general $L$. The results are presented in figure \ref{figure:fz0}. For large $L$ we find that $F(L T)$ always converges to a certain value, independent of the dimensionality of the theory. For large $L$ and $d$, $F(z_0)$ approaches $1/4$. For large $d$, $F(LT)$ jumps quicker for even intermediate values of $LT$ to the plateau $\sim 1/4$. In general in the large $d$ regime the integrals converge to a ceratin curve reflecting the narrowing of the gravitational potential around the black hole horizon. This is clearly evident in the computation of the entanglement entropy with respect to the boundary distance $L$, presented in figure \ref{figure:eLr1}. As the dimension increases the entanglement converges fast to a certain form, which is described by a simpler function of $L$.

As a final remark we note that the entanglement entropy being an area of a codimension one surface depends stronger on $d$ than the Wilson loop. Nevertheless, for large $d$, its dependence on the length strip $L$ still approaches a certain form and further increase of $d$ does not cause further change. This is the regime that the gravitational potential is strongly localized in the near horizon.  To make a quantitative comparison, the entanglement entropy of a strip of length $L\simeq 2$ gets modified by $30\%$ compared to the same strip, say for example in $d=25$. This extrapolation can be considered as a well accepted one.


In the next section we generalize our computation to space-like surfaces of codimension $q$,  which include the entangling ones for $q=d-1$ and we express the expectation values in closed form  of infinite converging series in terms of $q$ and $d$. We also take the large $d$ limit to show analytically that the series converge to certain values and that $F(z_h)\simeq1/4$.

\subsection{ Spacelike Surfaces of Codimension $q$} \label{section:qen}

In this section we generalize our results for space-like bulk surfaces of spatial codimension $d-q$. The boundary conditions are of a slab spatial volume of dimension $q$, separated along the $x$ direction. For the special case of $q=d-1$, we get the entangling surface. The area for the generic surface reads
\be\la{eesq}
 S=\cR^{q-1} \int d\s g_{xx}^{\ff{q-1}{2}}\sqrt{g_{xx}+g_{zz}z'(\s)^2}~,
\ee
where $\cR^{q-1}$ is the $(q-1)$-dimensional spatial volume of the dimensions transverse to the $x$ direction. This area will be proportional to the entanglement entropy in lower dimensional boundary theories, once we compactify the directions transverse to the slab. The first order equation of motion reads
\be
z'(\s)^2= \ff{g_{xx}\prt{g_{xx}^{q}-c^2}}{c^2 g_{zz}}~,
\ee
where the constant $c^2:=g_{xx}^{q}(z_0)$ with $z_0$ being the turning point of the surface. The boundary length of the surface is then given as
\be
L=2 \int  \sqrt{\ff{c^2 g_{zz}}{g_{xx}\prt{g_{xx}^{q}- c^2}}} dz~,
\ee
while the extremal area
\be
 S=\cR^{q-1}\int dz g_{xx}^{\ff{q-1}{2}}\sqrt{\ff{g_{zz} g_{xx}^q}{g_{xx}^q-c^2}}\,.
\ee
Applying our generic analysis on the background under study, we get
\be\la{leeq}
L=2 z_0 \int_0^{1} \ff{y^{q}}{\sqrt{\prt{1-w^d y^d} \prt{1- y^{2 q}}}} dy~.
\ee
and for the area
\be\la{seehy}
 S=2 \cR^{q-1} R^q \ff{1}{z_0^{q-1}}\int_{0}^{1} dy \ff{1}{y^{q} \sqrt{\prt{1-w^d y^d}\prt{1-y^{2 q}}}}~,
\ee
where $w:=\ff{z_0}{z_h}$  . The divergence is of the order of $y^{q}$ and we isolate the relevant term by adding and subtracting the term responsible for the divergence. To extract the large $d$ contributions, we write the area in the following form
\be\la{eearea}
S= \cR^{q-1} R^q \prtt{\ff{L}{z_0^{d-1}}+ \ff{2}{\prt{q-1} \e^{q-1}} +\ff{2}{z_0^{q-1}}\prt{ F(z_0)-\ff{1}{q-1}}}~,
\ee
where
\be\la{fz0q}
F(z_0)=  \int_0^{1} dy \ff{1}{y^{q}}\prt{\sqrt{\ff{1-y^{2q}}{1-w^d y^d}}-1}~.
\ee
The expressions cannot be obtained in closed form in terms of elementary functions. Instead they can be written in terms of an infinite converging series. The length is equal to
\be
L=2 z_0 \ff{\Gamma\prt{n+\ff{1}{2}}}{\G\prt{n+1}}\ff{\G\prt{\ff{d n+1}{2 q}+\ff{1}{2}}}{\prt{1+d n}\G\prt{1+\ff{d n}{2 q}}} \prt{\ff{z_0}{z_h}}^{n d}\,,
\ee
where we have applied the binomial series once and then we have integrated the expression.
Then the area requires a more careful consideration, due to infinities. Let us concentrate on the term $F(z_0)$ given by \eq{fz0q}. By applying the binomial expansion we get
\be
F(z_0)= \int_0^{1} dy \prtt{\ff{1}{y^{q}}\prt{\sqrt{1-y^{2q}}-1}+\sqrt{1-y^{2q}}\sum_{n=1}^\infty c_n w^{n d} y^{n d-q}}
\ee
where we have separated the first term of the sum in order to group the potential infinities. The integrals then can be performed analytically to get
\be\la{fz00}
F(z_0)=\ff{1}{q-1}+\ff{\sqrt{\pi}\G\prt{\ff{1}{2} \prt{\ff{1}{q}-1}}}{2 \Gamma\prt{\ff{1}{2 q}}} +
\ff{\sqrt{\pi}}{4 q} \sum_{n=1}^\infty c_n\ff{\G\prt{\ff{1-q+d n}{2 q}}}{\G\prt{\ff{1+d n}{2 q}+1}} w^{d n}~,
\ee
for $q\neq1$. Therefore the extremal surface \eq{eearea} of spatial codimension $d-q$  has been written in terms of infinite converging series.

Notice that the entangling surface is a special case of this expression for $q=d-1$. It is interesting to apply for this case the limit $d\rightarrow \infty$ in \eq{fz00}. By applying the limits carefully and considering the leading contributions of the sum for low values of $n$ and $w\rightarrow 1$, which is equivalent to $z_0\rightarrow z_h$, we obtain
\be
F(z_0)\simeq \ff{1}{4} w^d~,~~\mbox{ at large $d$},
\ee
which reproduces the right result $F(z_h)\simeq 1/4$, following the earlier numerical treatment of the figure \ref{figure:fzh} and the analytical one with the hypergeometric functions \eq{fres}.


As an extra remark, we note that our current analysis focuses on the study of entangling regions with slab geometry which is interesting. The large-$d$ analysis of different type of surfaces is interesting in its own right. For different shapes of  entangling surfaces the asymptotic form of the observables is qualitatively different for even and odd $d$.  This is evident even in the zero temperature limit where the entanglement entropy of a sphere of radius $L$ takes the form \cite{Ryu:2006ef,Gushterov:2017vnr}
\be\la{disk1}
S= \ff{R^{d-1}Vol(S^{d-2})}{4 G_N^{(d+1)}}\prt{\sum_{i=1}^{(d-k)/2} p_i \prt{\ff{L}{\e}}^{d-2 i} + p_{0} +\tilde{p}_{0} \log\prt{\ff{L}{\e}}+\ldots}~,
\ee
where $\tilde{p}_{0}$ is zero when $d$ is odd, and $k=2(1)$  for even (odd) $d$. 
Assuming that the asymptotic expansion commutes with the large $d$ expansion, the leading polynomial terms of \eq{disk1} admit a common limit for large $d$, independently of the parity. However there are terms like the logarithmic one, that  appear only in even or odd dimensions. For generic entangling surfaces, in order to keep track of all terms at large $d$ that depend on the parity of $d$, we can express the corresponding surface integral in terms of infinite converging series, obtaining an expression like \eq{fz00}. Notice that essentially we need to take into account subleading terms in the large $d$ expansion. The infinite series provide information on how the large $d$-limit is sensitive on the parity of $d$. It would be interesting to investigate these issues further as part of future work.

\section{Conclusions}

In this paper we have considered minimal surfaces of arbitrary codimension in $AdS$ spaces of arbitrary dimensionality. Our analysis includes Wilson loops as well as entangling surfaces and the study of the monotonicity of the area theorem. We have considered the limit of large number of dimensions $d$, both analytically and numerically. We rely on the fact that the bulk geometry is well defined at any dimension and we assume that the holographic dictionary between extremal surfaces and non-local operators is valid at any $d$. Additionally, we show that the large $d$ expansion in our holographic study can be used both as a conceptual and a computational tool, since all the computational limits are smooth. We find that the large $d$ analysis captures all the qualitative holographic features of the low $d$ analysis. Moreover, the quantitative analysis of the observables is in good agreement with the low $d$ analysis, despite the desirable simplifications in this limit. We also observe convergence of our results to certain values, hinting at certain universal behaviors in this regime. This partly relies on the fact that for large $d$, the gravitational potential becomes extremely steep, localized in the near horizon regime with a steep gradient, which tends to separate the dynamics of the near horizon regime and the rest of the space. At the same time, it preserves the generic and holographic characteristics of the gravitational potential since a horizon is present.

Moreover, all the expectation values of the observables under study show rapid convergence to certain values as $d$ increases. Therefore, in practice, the large $d$ limit is a limit of finite intermediate-low dimensions. In particular, the convergence of the observables to their large $d$ limit values has been already achieved for dimensions around $d=10$ to $d=20$.   The extrapolation of the large $d$ limit  to low dimensions shows a good quantitative agreement with the numerical analysis on the observables.  In particular, the extrapolation leads to precision that is a few percent off to about 30 percent off at most, as compared to the exact numerical analysis. To some extend this reminds the behavior of certain observables and their dependence on large $N$ limit, for example \cite{Bali:2013kia}, although the mechanisms involved are very are different. It is worthy to mention that similar quantitative precision and dependency on the dimensionality have been observed in different holographic studies, for example the phase transition of the mutual information at zero temperature  \cite{Colin-Ellerin:2019vst}. This suggests that the large $d$ expansion can in general be considered as a useful tool in holography.

In addition, we were able to study analytically  the area theorem and to demonstrate that it is violated in the large $d$ limit in our theories. The violation relies on a function that converges at large $d$ to a certain value that can be computed analytically. The violation itself is not a surprise since the theory is in the thermal deconfining phase with broken Lorentz invariance and the monotonicity theorems are known to be difficult to satisfy in such cases \cite{Gushterov:2017vnr,Chu:2019uoh}. However, the fact that the computation in the limit of large dimensions is tractable analytically and conclusive, consists a significant advantage. This suggests that the large $d$-limit as an invaluable tool for the search, study, and (in)validation  of the RG monotonicity theorems.

We have also generalized our analysis to arbitrary codimension Wilson loop surfaces and space-like surfaces expressing our results in closed form of infinite converging series that depend on the dimensionality of the space and the codimension of the surfaces. We have taken there the large $d$ limit to show how these expressions simplify.

Our studies can be extended in several ways. One could perform explicit computations of  several other observables related to the minimization of the string and brane actions.  We expect that the generic features found here will carry on to all such observables in a similar way. Such computations would further establish the use of the large $d$ limit as a holographic tool. In the same spirit one could also try to construct the Einstein equations from Wilson loop or entanglement entropy relations and study the bulk reconstruction in the large $d$ limit. It would be also very interesting to examine the effect of the large $d$ limit in (marginally) deformed theories,  with broken global symmetries. For example, the anisotropic axion backreacted theories of \cite{Giataganas:2017koz,Mateos:2011ix} and their observables \cite{Giataganas:2012zy} could be considerably simplified at large dimensions. In this limit the dual backgrounds could be derived analytically and perturbatively irrespective of the magnitude of the anisotropic deformation, which otherwise is not possible.

\textbf{Acknowledgments:}
We would like to thank N. Irges for useful discussions. The research work of D.G. is supported by  Ministry of Science and Technology of Taiwan (MOST) by the Young Scholar Columbus Fellowship grant 110-2636-M-110-007. The research work of N.P. is supported by the Hellenic Foundation for Research and Innovation (H.F.R.I.) and the General Secretariat for Research and Technology (G.S.R.T.), under grant agreement No 2344.


\bibliographystyle{JRmod}


\begin{thebibliography}{10}

\bibitem{Aharony:1999ti}
O.~Aharony, S.~S. Gubser, J.~M. Maldacena, H.~Ooguri and Y.~Oz, \emph{Large n
  field theories, string theory and gravity}, {\emph{Phys. Rept.} {\bfseries
  323} (2000) 183--386},
  [\href{https://arxiv.org/abs/hep-th/9905111}{{\ttfamily hep-th/9905111}}].

\bibitem{CasalderreySolana:2011us}
J.~Casalderrey-Solana, H.~Liu, D.~Mateos, K.~Rajagopal and U.~A. Wiedemann,
  \emph{{Gauge/String Duality, Hot QCD and Heavy Ion Collisions}},
  \href{https://arxiv.org/abs/1101.0618}{{\ttfamily 1101.0618}}.

\bibitem{cardyc}
J.~L. Cardy, \emph{Is there a c-theorem in four dimensions?},
  \href{https://doi.org/https://doi.org/10.1016/0370-2693(88)90054-8}{\emph{Physics
  Letters B} {\bfseries 215} (1988) 749--752}.

\bibitem{Giombi:2014xxa}
S.~Giombi and I.~R. Klebanov, \emph{{Interpolating between $a$ and $F$}},
  \href{https://doi.org/10.1007/JHEP03(2015)117}{\emph{JHEP} {\bfseries 03}
  (2015) 117}, [\href{https://arxiv.org/abs/1409.1937}{{\ttfamily 1409.1937}}].

\bibitem{Ryu:2006ef}
S.~Ryu and T.~Takayanagi, \emph{{Aspects of Holographic Entanglement Entropy}},
  \href{https://doi.org/10.1088/1126-6708/2006/08/045}{\emph{JHEP} {\bfseries
  08} (2006) 045}, [\href{https://arxiv.org/abs/hep-th/0605073}{{\ttfamily
  hep-th/0605073}}].

\bibitem{Myers:2010tj}
R.~C. Myers and A.~Sinha, \emph{{Holographic c-theorems in arbitrary
  dimensions}}, \href{https://doi.org/10.1007/JHEP01(2011)125}{\emph{JHEP}
  {\bfseries 01} (2011) 125},
  [\href{https://arxiv.org/abs/1011.5819}{{\ttfamily 1011.5819}}].

\bibitem{Chu:2019uoh}
C.-S. Chu and D.~Giataganas, \emph{{$c$-Theorem for Anisotropic RG Flows from
  Holographic Entanglement Entropy}},
  \href{https://doi.org/10.1103/PhysRevD.101.046007}{\emph{Phys. Rev. D}
  {\bfseries 101} (2020) 046007},
  [\href{https://arxiv.org/abs/1906.09620}{{\ttfamily 1906.09620}}].

\bibitem{Giataganas:2017koz}
D.~Giataganas, U.~G\"ursoy and J.~F. Pedraza, \emph{{Strongly-coupled
  anisotropic gauge theories and holography}},
  \href{https://doi.org/10.1103/PhysRevLett.121.121601}{\emph{Phys. Rev. Lett.}
  {\bfseries 121} (2018) 121601},
  [\href{https://arxiv.org/abs/1708.05691}{{\ttfamily 1708.05691}}].

\bibitem{Baggioli:2020cld}
M.~Baggioli and D.~Giataganas, \emph{{Detecting Topological Quantum Phase
  Transitions via the c-Function}},
  \href{https://doi.org/10.1103/PhysRevD.103.026009}{\emph{Phys. Rev. D}
  {\bfseries 103} (2021) 026009},
  [\href{https://arxiv.org/abs/2007.07273}{{\ttfamily 2007.07273}}].

\bibitem{Cremonini:2020rdx}
S.~Cremonini, L.~Li, K.~Ritchie and Y.~Tang, \emph{{Constraining
  nonrelativistic RG flows with holography}},
  \href{https://doi.org/10.1103/PhysRevD.103.046006}{\emph{Phys. Rev. D}
  {\bfseries 103} (2021) 046006},
  [\href{https://arxiv.org/abs/2006.10780}{{\ttfamily 2006.10780}}].

\bibitem{Hoyos:2021vhl}
C.~Hoyos, N.~Jokela, J.~M. Pen\'\i{}n, A.~V. Ramallo and J.~Tarr\'\i{}o,
  \emph{{Risking your NEC}},
  \href{https://arxiv.org/abs/2104.11749}{{\ttfamily 2104.11749}}.

\bibitem{Cartwright:2021hpv}
C.~Cartwright and M.~Kaminski, \emph{{Inverted c-functions in thermal states}},
   \href{https://arxiv.org/abs/2107.12409}{{\ttfamily 2107.12409}}.

\bibitem{Arefeva:2020uec}
I.~Y. Aref'eva, A.~Patrushev and P.~Slepov, \emph{{Holographic entanglement
  entropy in anisotropic background with confinement-deconfinement phase
  transition}}, \href{https://doi.org/10.1007/JHEP07(2020)043}{\emph{JHEP}
  {\bfseries 07} (2020) 043},
  [\href{https://arxiv.org/abs/2003.05847}{{\ttfamily 2003.05847}}].

\bibitem{Cremonini:2013ipa}
S.~Cremonini and X.~Dong, \emph{{Constraints on renormalization group flows
  from holographic entanglement entropy}},
  \href{https://arxiv.org/abs/1311.3307}{{\ttfamily 1311.3307}}.

\bibitem{Casini:2012ei}
H.~Casini and M.~Huerta, \emph{{On the RG running of the entanglement entropy
  of a circle}}, \href{https://doi.org/10.1103/PhysRevD.85.125016}{\emph{Phys.
  Rev. D} {\bfseries 85} (2012) 125016},
  [\href{https://arxiv.org/abs/1202.5650}{{\ttfamily 1202.5650}}].

\bibitem{Casini:2016udt}
H.~Casini, E.~Teste and G.~Torroba, \emph{{Relative entropy and the RG flow}},
  \href{https://doi.org/10.1007/JHEP03(2017)089}{\emph{JHEP} {\bfseries 03}
  (2017) 089}, [\href{https://arxiv.org/abs/1611.00016}{{\ttfamily
  1611.00016}}].

\bibitem{Emparan:2013moa}
R.~Emparan, R.~Suzuki and K.~Tanabe, \emph{{The large D limit of General
  Relativity}}, \href{https://doi.org/10.1007/JHEP06(2013)009}{\emph{JHEP}
  {\bfseries 06} (2013) 009},
  [\href{https://arxiv.org/abs/1302.6382}{{\ttfamily 1302.6382}}].

\bibitem{Emparan:2013xia}
R.~Emparan, D.~Grumiller and K.~Tanabe, \emph{{Large-D gravity and low-D
  strings}}, \href{https://doi.org/10.1103/PhysRevLett.110.251102}{\emph{Phys.
  Rev. Lett.} {\bfseries 110} (2013) 251102},
  [\href{https://arxiv.org/abs/1303.1995}{{\ttfamily 1303.1995}}].

\bibitem{Emparan:2014cia}
R.~Emparan and K.~Tanabe, \emph{{Universal quasinormal modes of large D black
  holes}}, \href{https://doi.org/10.1103/PhysRevD.89.064028}{\emph{Phys. Rev.
  D} {\bfseries 89} (2014) 064028},
  [\href{https://arxiv.org/abs/1401.1957}{{\ttfamily 1401.1957}}].

\bibitem{Dandekar:2016fvw}
Y.~Dandekar, A.~De, S.~Mazumdar, S.~Minwalla and A.~Saha, \emph{{The large D
  black hole Membrane Paradigm at first subleading order}},
  \href{https://doi.org/10.1007/JHEP12(2016)113}{\emph{JHEP} {\bfseries 12}
  (2016) 113}, [\href{https://arxiv.org/abs/1607.06475}{{\ttfamily
  1607.06475}}].

\bibitem{Colin-Ellerin:2019vst}
S.~Colin-Ellerin, V.~E. Hubeny, B.~E. Niehoff and J.~Sorce, \emph{{Large-$d$
  phase transitions in holographic mutual information}},
  \href{https://doi.org/10.1007/JHEP04(2020)173}{\emph{JHEP} {\bfseries 04}
  (2020) 173}, [\href{https://arxiv.org/abs/1911.06339}{{\ttfamily
  1911.06339}}].

\bibitem{Andrade:2015hpa}
T.~Andrade, S.~A. Gentle and B.~Withers, \emph{{Drude in D major}},
  \href{https://doi.org/10.1007/JHEP06(2016)134}{\emph{JHEP} {\bfseries 06}
  (2016) 134}, [\href{https://arxiv.org/abs/1512.06263}{{\ttfamily
  1512.06263}}].


\bibitem{Garcia-Garcia:2015emb}
A.~M.~Garc\'\i{}a-Garc\'\i{}a and A.~Romero-Berm\'udez,
``Conductivity and entanglement entropy of high dimensional holographic superconductors,''
JHEP \textbf{09} (2015), 033, [\href{https://arxiv.org/abs/1502.03616}{{\ttfamily
  1502.03616}}].



\bibitem{Rozali:2017bll}
M.~Rozali, E.~Sabag and A.~Yarom, \emph{{Holographic Turbulence in a Large
  Number of Dimensions}},
  \href{https://doi.org/10.1007/JHEP04(2018)065}{\emph{JHEP} {\bfseries 04}
  (2018) 065}, [\href{https://arxiv.org/abs/1707.08973}{{\ttfamily
  1707.08973}}].

\bibitem{Casalderrey-Solana:2018uag}
J.~Casalderrey-Solana, C.~P. Herzog and B.~Meiring, \emph{{Holographic Bjorken
  Flow at Large-$D$}},
  \href{https://doi.org/10.1007/JHEP01(2019)181}{\emph{JHEP} {\bfseries 01}
  (2019) 181}, [\href{https://arxiv.org/abs/1810.02314}{{\ttfamily
  1810.02314}}].

\bibitem{Andrade:2018zeb}
T.~Andrade, C.~Pantelidou and B.~Withers, \emph{{Large D holography with metric
  deformations}}, \href{https://doi.org/10.1007/JHEP09(2018)138}{\emph{JHEP}
  {\bfseries 09} (2018) 138},
  [\href{https://arxiv.org/abs/1806.00306}{{\ttfamily 1806.00306}}].

\bibitem{Brandhuber:1998bs}
A.~Brandhuber, N.~Itzhaki, J.~Sonnenschein and S.~Yankielowicz, \emph{{Wilson
  loops in the large N limit at finite temperature}},
  \href{https://doi.org/10.1016/S0370-2693(98)00730-8}{\emph{Phys. Lett.}
  {\bfseries B434} (1998) 36--40},
  [\href{https://arxiv.org/abs/hep-th/9803137}{{\ttfamily hep-th/9803137}}].

\bibitem{Fischler:2012ca}
W.~Fischler and S.~Kundu, \emph{{Strongly Coupled Gauge Theories: High and Low
  Temperature Behavior of Non-local Observables}},
  \href{https://doi.org/10.1007/JHEP05(2013)098}{\emph{JHEP} {\bfseries 05}
  (2013) 098}, [\href{https://arxiv.org/abs/1212.2643}{{\ttfamily 1212.2643}}].

\bibitem{Erdmenger:2017pfh}
J.~Erdmenger and N.~Miekley, \emph{{Non-local observables at finite temperature
  in AdS/CFT}}, \href{https://doi.org/10.1007/JHEP03(2018)034}{\emph{JHEP}
  {\bfseries 03} (2018) 034},
  [\href{https://arxiv.org/abs/1709.07016}{{\ttfamily 1709.07016}}].

\bibitem{Horowitz:1998ha}
G.~T. Horowitz and R.~C. Myers, \emph{{The AdS / CFT correspondence and a new
  positive energy conjecture for general relativity}},
  \href{https://doi.org/10.1103/PhysRevD.59.026005}{\emph{Phys. Rev. D}
  {\bfseries 59} (1998) 026005},
  [\href{https://arxiv.org/abs/hep-th/9808079}{{\ttfamily hep-th/9808079}}].

\bibitem{Hubeny:2009rc}
V.~E. Hubeny, D.~Marolf and M.~Rangamani, \emph{{Hawking radiation from AdS
  black holes}},
  \href{https://doi.org/10.1088/0264-9381/27/9/095018}{\emph{Class. Quant.
  Grav.} {\bfseries 27} (2010) 095018},
  [\href{https://arxiv.org/abs/0911.4144}{{\ttfamily 0911.4144}}].

\bibitem{Drukker:1999zq}
N.~Drukker, D.~J. Gross and H.~Ooguri, \emph{{Wilson loops and minimal
  surfaces}}, \href{https://doi.org/10.1103/PhysRevD.60.125006}{\emph{Phys.
  Rev.} {\bfseries D60} (1999) 125006},
  [\href{https://arxiv.org/abs/hep-th/9904191}{{\ttfamily hep-th/9904191}}].

\bibitem{Chu:2008xg}
C.-S. Chu and D.~Giataganas, \emph{{UV-divergences of Wilson Loops for
  Gauge/Gravity Duality}},
  \href{https://doi.org/10.1088/1126-6708/2008/12/103}{\emph{JHEP} {\bfseries
  0812} (2008) 103}, [\href{https://arxiv.org/abs/0810.5729}{{\ttfamily
  0810.5729}}].

\bibitem{Gushterov:2017vnr}
N.~I. Gushterov, A.~O'Bannon and R.~Rodgers, \emph{{On Holographic Entanglement
  Density}}, \href{https://doi.org/10.1007/JHEP10(2017)137}{\emph{JHEP}
  {\bfseries 10} (2017) 137},
  [\href{https://arxiv.org/abs/1708.09376}{{\ttfamily 1708.09376}}].

\bibitem{Bali:2013kia}
G.~S. Bali, F.~Bursa, L.~Castagnini, S.~Collins, L.~Del~Debbio, B.~Lucini
  et~al., \emph{{Mesons in large-N QCD}},
  \href{https://doi.org/10.1007/JHEP06(2013)071}{\emph{JHEP} {\bfseries 06}
  (2013) 071}, [\href{https://arxiv.org/abs/1304.4437}{{\ttfamily 1304.4437}}].

\bibitem{Mateos:2011ix}
D.~Mateos and D.~Trancanelli, \emph{{The anisotropic N=4 super Yang-Mills
  plasma and its instabilities}},
  \href{https://doi.org/10.1103/PhysRevLett.107.101601}{\emph{Phys.Rev.Lett.}
  {\bfseries 107} (2011) 101601},
  [\href{https://arxiv.org/abs/1105.3472}{{\ttfamily 1105.3472}}].

\bibitem{Giataganas:2012zy}
D.~Giataganas, \emph{{Probing strongly coupled anisotropic plasma}},
  \href{https://doi.org/10.1007/JHEP07(2012)031}{\emph{JHEP} {\bfseries 1207}
  (2012) 031}, [\href{https://arxiv.org/abs/1202.4436}{{\ttfamily 1202.4436}}].

\end{thebibliography}

\end{document}